\documentclass[aip,pop,reprint,groupedaddress]{revtex4-1}

\draft 

\usepackage{graphicx}
\usepackage{siunitx}
\usepackage{float}

\begin{document}

\preprint{LA-UR-16-28903}

\title{Semi-analytic model of plasma-jet-driven magneto-inertial fusion} 



\author{Samuel J. Langendorf}
\email[]{samuel.langendorf@lanl.gov}

\author{Scott C. Hsu}
\email[]{scotthsu@lanl.gov}
\affiliation{Physics Division, Los Alamos National Laboratory, Los Alamos, NM  87545}


\date{\today}

\begin{abstract}
A semi-analytic model for plasma-jet-driven magneto-inertial fusion is presented. Compressions of a magnetized plasma target by a spherically imploding plasma liner are calculated in one dimension (1D), accounting for compressible hydrodynamics and ionization of the liner material, energy losses due to conduction and radiation, fusion burn and alpha deposition, separate ion and electron temperatures in the target, magnetic pressure, and fuel burn-up. Results show 1D gains of 3--30 at spherical convergence ratio~$<\ 15$ and 20--40~MJ of liner energy, for cases in which the liner thickness is 1 cm and the initial radius of a preheated magnetized target is 4 cm. Some exploration of parameter space and physics settings is presented. The yields observed suggest that there is a possibility of igniting additional dense fuel layers to reach high gain. 
\end{abstract}


\maketitle 

\section{Introduction}

Magneto-inertial fusion (MIF), aka magnetized target fusion (MTF), \cite{lindemuth83,kirkpatrick95,thio08,lindemuth09,lindemuth15} is an approach to laboratory fusion that operates on time scales and power levels in between the extremes of inertial confinement fusion (ICF) and magnetic confinement fusion (MCF)\@. The addition of a magnetic field in the plasma ``target'' to be compressed potentially aids in igniting the target via inhibition of thermal conduction across the field as well as through enhanced retention of the energy of charged particles released in fusion reactions. The ignition criterion for an MIF system has been shown to be based on the achieved product $BR$ of compressed magnetic field strength and target radius,\cite{basko00} thereby differing fundamentally from the areal-density-based criterion for ICF ignition and opening up the possibility of achieving ignition at lower driver power densities and
velocities (in the 10--100-km/s range). Recently, the magnetically driven, cylindrical Magnetized-Liner-Inertial-Fusion (MagLIF) concept\cite{slutz10} has yielded very promising experimental results, demonstrating fusion-relevant temperatures and $BR$ values.\cite{gomez14,schmit14}

Plasma-jet-driven MIF, or PJMIF,\cite{thio99,hsu09,hsu12} is an MIF architecture possessing the desirable attribute of ``standoff,'' i.e., a way to keep the compressed plasma physically distant from the facility first wall and avoid repetitive driver-hardware destruction. The concept is to form a spherically imploding plasma liner from the convergence of a large number of discrete supersonic plasma jets, and use the assembled liner to compress a magnetized DT plasma target. The target may include an outer layer of cooler, denser DT (``afterburner'')\cite{lindemuth1991promise} that may heat and burn from energy deposition by alpha particles born in the inner portion of the target. Attaining standoff from the fusion plasma allows heat loads to be spread over larger areas, and thus damage/activation of the facility walls to be lessened and/or mitigated. Plasma jets may be supplied by contoured-gap coaxial plasma guns,\cite{witherspoon09} which could potentially achieve efficiencies of $> 50$\% and enable a fusion reactor at relatively modest gains $\lesssim$ 20 (i.e., fusion energy divided by the liner- and target-formation energies).

There have been several works in recent years relating to PJMIF, either directly or indirectly. One-dimensional (1D) simulations of desired PJMIF configurations\cite{knapp14} indicated the possibility for suitable levels of fusion gain ($\approx$ 30) with cm-thick plasma liners in the 40--100~MJ kinetic-energy range. Results based on an 1D analytical model of MIF implosions\cite{lindemuth15} also found promising energy-gain results, especially in spherical geometry. Analytical works\cite{parks08,cassibry09} have studied the hydrodynamic scaling properties of imploding plasma liners, and 1D hydrodynamic simulations of plasma-liner implosions\cite{awe11,davis12} and target compression\cite{samulyak10,santarius12,kim12} have also been performed, as have 3D simulations of plasma-jet merging and plasma-liner formation.\cite{cassibry12,kim13,cassibry13} Experimentally, PJMIF-relevant studies have been conducted on plasma-jet propagation,\cite{case10,hsu12pop} as well as oblique two-jet\cite{merritt13,merritt14} and three-jet merging.\cite{case13,messer13} Experiments over the next few years are planned to demonstrate and study plasma-liner formation with up to 60 plasma jets. 

Semi-analytic models have been productively used in MIF to explore interactions of the diverse physics. The seminal models of Lindemuth and Kirkpatrick\cite{lindemuth83,kirkpatrick95} are representative of generalized MIF implosions, but treat the liner hydrodynamics as the motion of an infinitely thin interface. The liner hydrodynamics may be important for PJMIF due to the extended length of available plasma jets. In recent years, McBride and Slutz's semi-analytic model of MagLIF\cite{mcbride15,mcbride16} treated the liner hydrodynamics by dividing the liner up into separate shells. The model was able to achieve reasonable agreement with 1D LASNEX simulations and provided insight into the physics. Inspired by both these models, we formulate a 1D, spherical-geometry, semi-analytic model for exploring PJMIF and provide an accompanying numerical code\cite{code} that typically executes in one minute or so on a personal computer. One can readily use the model to scan a large PJMIF parameter space, allowing interested researchers to identify attractive PJMIF configurations, key issues and obstacles, and PJMIF-development technology needs. Because a target fusion gain greater than unity is likely required to attain significant burn in an afterburner, in this paper we first focus on cases without an afterburner and defer detailed studies of cases with an afterburner for future work.

The paper is organized as follows.  Section~\ref{sec:model} presents the details of the model.  Section~\ref{sec:results}
presents results and some investigations of parameter space and physics settings. Conclusions are presented in Sec.~IV.

\section{Model formulation}
\label{sec:model}

We consider the PJMIF implosion in two phases: liner formation and convergence, and target compression.  In the liner formation and convergence phase, discrete jets propagate from their injection radius $R_{ji}$ towards the chamber center, until they encounter each other prior to reaching the center. We refer to this region of the encounter as the ``merging radius'' $R_m$. Once the jets have merged, the liner is assumed to converge as a spherical shell towards the center of the chamber.  The target is assumed to have been formed or be in the process of being formed at a particular initial radius $R_{T0}$; the target formation could occur by merging of injected compact toroids, or by stagnating unmagnetized jets and driving a magnetizing current as has been proposed in the laser beat-wave-magnetization approach.\cite{welch2012simulations,welch2014particle} When the converging liner reaches $R_{T0}$, we consider this to be the beginning of target compression.  The liner compresses the target, and is decelerated and ultimately stagnated and repulsed by the increasing pressure of the target. If the spherical compression is adiabatic, target pressure rises as convergence ratio $CR^5$, where $CR = R_{T0} / R_T$. 

The target compression phase determines fusion yield and gain, so we first consider the target compression phase to determine requirements for the merged liner, and then work backwards through the convergence and formation steps in PJMIF to investigate how such a liner might be formed.

\subsection{Target compression}

The MIF target compression process spans a diverse set of physics, including compressible hydrodynamics of the liner and target, ionization, radiation, magnetic pressure and advection, heat conduction, fusion burn, and charged particle energy deposition, all of which we describe in this section.  Many treatments have been applied by previous authors; where we deviate, we describe the motivation and implementation. These physics areas are certainly not exhaustive even in 1D; for example we do not study liner-target mix effects, though we do estimate the collisional interpenetration depths to be small in comparison to the target. We consider only 1D and do not attempt to account for instabilities, such as Rayleigh-Taylor instabilities that may occur at the decelerating liner-target interface.

\subsubsection{Hydrodynamics}

Previous analytical treatments have modeled the liner and target hydrodynamics assuming polytropic equations of state (EOS). Work in the MagLIF model showed that treating the liner as a single fluid element with a polytropic index $\gamma = 5/3$ yielded overly optimistic compression results. The issue was resolved by considering the liner as a series of cylindrical shells, effectively implementing Lagrangian hydrodynamics. A similar effect was achieved previously in the Lindemuth and Kirkpatrick models\cite{lindemuth83, kirkpatrick1979overview} by using a greater liner $\gamma$ of $2.5$. Because $Pv^\gamma$ is considered constant in polytropic EOS, increasing $\gamma$ increases the liner pressure and self-work during convergence and halts the implosion earlier. 

The details of the hydrodynamics may be important for PJMIF liners, as they may be thick and have longer stagnation times than other concepts, and thus it is necessary to treat the shock propagating back through the thick liner at stagnation. Therefore we opt for the Lagrangian hydrodynamics approach in the liner and discretize it into a series of spherical shells of equal mass. A quadratic artificial viscosity $q$ is included to cope with shocks. The acceleration of the liner zone interfaces is calculated for the interior zone interfaces as
\begin{equation}
a_{zi,1:N-1} = (p_{1:N} - p_{0:N-1} + q_{1:N} - q_{0:N-1}) * A_{1:N-1} / m_{1:N-1} ,
\end{equation}
in which $N$ is the number of zones, $p$ is the pressure, $q$ is the artificial viscosity, $A$ is the area of the spherical shell interface, and $m$ is the mass assigned to the spherical shell interface. For these interior zones, $m$ is assigned as $m_L / N$. The acceleration of the liner-target interface uses the target pressure $p_T$,
\begin{equation}
a_{zi,0} = (p_T - p_{0} - q_{0}) * A_{0} / m_{0} ,
\end{equation}
\begin{equation}
p_T = p_{T,i} + p_{T,e} + p_{T,r} + p_{T,B} ,
\end{equation}
in which subscripts $i$,~$e$,~$r$~and~$B$ represent ion, electron, radiation, and magnetic quantities and the assigned $m_0 = m_L / 2N$.  For the exterior liner zone, a user-input vacuum pressure $p_{vac}$ is included, and the acceleration is calculated as 
\begin{equation}
a_{zi,N} = (p_{N} - p_{vac} + q_{N}) * A_{N} / m_{N} ,
\end{equation}
\begin{equation}
m_N = \frac{m_L}{2N} .
\end{equation}
We assume that the target has been preheated to around 100 eV such that fusion temperatures can be obtained at modest spherical convergence $\leq 15$. The target has a high sound speed, so we assume that it is isobaric and consider it as a single fluid element. We assume $T_i = T_e = T_r$ throughout the liner, and allow for the possibility of $T_i \neq T_e \neq T_r$ in the target, where it has been indicated that high temperatures lessen the collisional coupling between the ions and electrons on the implosion time scale.\cite{santarius12}

With zone accelerations computed, zone interface velocities and positions can be updated, and zone volumes, densities, and energy transfer rates are then computed. The energy transfer rate terms are the core of the model and represent the physics of interest, including compression work, heat conduction, radiation, alpha deposition, etc.  They are used to update the internal energy per unit mass of the target and liner as
\begin{equation}
e_{i,T,t} = e_{i,T,t-1} + \frac{\Delta t}{m_T} \left( P_{hydro,i} + \frac{P_{\alpha}}{2} - P_{hc,i}  + P_{ie}\right ) ,
\end{equation}
\begin{equation}
e_{e,T,t} = e_{e,T,t-1} + \frac{\Delta t}{m_T} \left( P_{hydro,e} + \frac{P_{\alpha}}{2} - P_{hc,e} - P_{ie}  + P_{er}\right ) ,
\end{equation}
\begin{equation}
e_{r,T,t} = e_{r,T,t-1} + \frac{\Delta t}{m_T} \left( P_{hydro,r} - P_{er} - P_{rad} \right ) ,
\end{equation}
\begin{equation}
e_{int,L,j,t} = e_{int,L,j,t-1} + \frac{\Delta t}{m_{L,j}}   \left( P_{hydro,j} + P_{hc,j} + P_{rad,j} \right ) ,
\end{equation}
in which subscript $j$ indicates a particular zone of the liner. The assorted power terms $P$ include hydrodynamic work $P_{hydro}$, alpha particle energy deposition $P_{\alpha}$, heat conduction $P_{hc}$, heat transfer due to radiation $P_{rad}$, ion-electron-equilibration $P_{ie}$ and radiation-matter coupling $P_{er}$, and subscripts $i$, $e$, and $r$ refer to ion, electron, and radiation fluids.  The amount of hydrodynamic compression / expansion work in the liner is found directly from the updated zone interface positions and mass densities as
\begin{equation}
\frac{\Delta t}{m_{L,j}}P_{hydro,j} = p_j \left(\frac{1}{\rho_{j,t}} - \frac{1}{\rho_{j,t-1}} \right),
\end{equation}
and the target hydrodynamic terms are dependent on the respective target pressures, as
\begin{equation}
P_{hydro,(i,e,r)} = p_{T,(i,e,r)} A_0 u_{zi,0},
\end{equation}
in which $u_{zi,0}$ is the velocity of the liner-target interface.

\subsubsection{Ionization}

\begin{figure}[!htb]
\begin{center}
\includegraphics[width=3.34in]{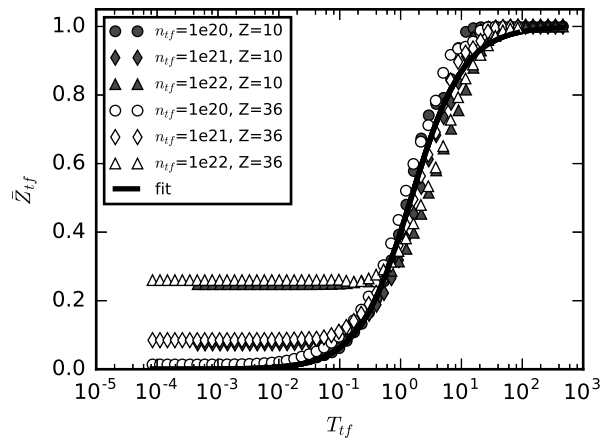}
\caption{Comparison between ionization fit and calculations from the Prism PROPACEOS code.\cite{macfarlane06}}
\label{fig:iz}
\end{center}
\end{figure}

We assume a Thomas-Fermi-like scaling of the liner mean charge $\bar Z$,
\begin{equation}
\bar Z_{tf} = \frac{\bar Z}{Z},
\label{eqn:zbar_tf}
\end{equation}
\begin{equation}
T_{tf} = \frac{T}{Z^{4/3}},
\end{equation}
where $Z$ is the atomic number of the liner species. We assume that for all liner species the mean charge is a function of reduced temperature only,
\begin{equation}
\bar Z_{tf} = \frac{1}{1 + 2/(T_{tf}^{0.85})}.
\label{eqn:zbar}
\end{equation}
As shown in Fig.~\ref{fig:iz}, this formula loosely approximates the results of local-thermodynamic-equilibrium (LTE) ionization calculations across a range of ion species, density, and temperatures from $10^{19}$--$10^{23}$~cm$^{-3}$ and 1--10000~eV\@. The electron and ion internal energy are specified as those of ion and electron fluids at the specified ionization per unit ion mass,
\begin{equation}
e_e = \bar Z \frac{3 N_A} {2 \mu} kT_e ,
\end{equation}
\begin{equation}
e_i = \frac{3 N_A} {2 \mu} kT_i ,
\end{equation}
in which $\mu$ is the ratio of ion mass to proton mass.

\subsubsection{Magnetic field}
\label{sec:mag}

An azimuthal magnetic flux $\Phi$ is assumed in the spherical target, generating an average magnetic flux density and pressure given by:
\begin{equation}
\bar B = \frac{\Phi}{\frac{1}{2} \pi R_T^2},
\end{equation}
\begin{equation}
p_{T,B} = \frac{\bar B^2}{2 \mu_0}.
\end{equation}
This value of magnetic energy density is used in the model for evolving energy balance, but for calculating transport, one might expect the magnetic field strength in the gradient layer at the target exterior to be somewhat lower than the flux-averaged field strength (though advection could change this). For the purposes of the present model, we assume a magnetic field profile due to an axial current in the sphere, and compute an effective reduced field strength in the gradient layer. The reduction is a function of the assumed gradient length scale, which we define with the symbol $\delta_1$. The reduced field strength is fit by
\begin{equation}
\frac{B_{eff}}{\bar B} = 0.18 \delta_1^2   +  0.41 \delta_1^2   +  0.41 .
\end{equation}
Lindemuth's model\cite{lindemuth15} considers magnetic flux loss from the target by resistive dissipation and the Nernst effect (flux transport arising due to temperature gradients). Since the target must be hot and timescales considered in PJMIF are short, we neglect resistive diffusion but consider the Nernst effect. Lindemuth includes Nernst flux diffusion using the Braginskii coefficient $\beta_\wedge^{uT}$, which in the magnetized limit decreases as $x_e^{-1}$ where $x_e$ is the target electron Hall parameter. We follow a similar approach, using the Epperlein-Heines\cite{epperlein1986plasma} formula for $\beta_\wedge^{uT}$ and a gradient-based expression for the resulting flux transport,
\begin{equation}
\beta_\wedge^{uT} = \frac{1.5 x_e^2 + 2.54 x_e}{x_e^3 + 7.09 x_e^2 + 3.27 x_e + 2.87} ,
\end{equation}
\begin{equation}
\dot \phi = -(2+\pi)R_T \frac{\beta_\wedge^{uT}}{e} \frac{T_{T,e} - T_{L,0}}{\delta_1 R_T} .
\end{equation}

The above analysis computes a flux loss, but may not be sufficient to fully represent the Nernst effect. A detailed analysis in cylindrical geometry by Velikovich\cite{velikovich2015magnetic} finds that Nernst flux transport is coupled with heat conduction and fluid advection in the target. Two main mechanisms are discussed: firstly, the cooling of the target exterior by heat conduction drives advection in the target, bringing material from the hot interior towards the cooler exterior to establish an isobaric condition. This advection carries magnetic flux and thermal energy towards the exterior of the target where it may be lost to the liner. Secondly, as magnetization and $x_e$ is increased, the thermal-gradient length scale becomes smaller, thus partially offsetting the inhibition of thermal transport provided by the increased magnetization. The results indicate that while magnetic flux losses decrease as $x_e^{-1}$, magnetized heat conduction losses may also only decrease as a Bohm-like $x_e^{-1}$ rather than the classical $x_e^{-2}$ scaling. We include an option to simulate this effect in the heat conduction by use of a modified thermal conductivity.

\subsubsection{Heat conduction}
\label{sec:therm}

We use a Braginskii-type\cite{braginskii1965transport} transport model for the thermal conductivity of the target, which incorporates reduction in heat conduction by the magnetic field. We use expressions for the Braginskii transport coefficients as given by Epperlein-Haines,\cite{epperlein1986plasma} which correct some known inaccuracies of the Braginskii treatment (as compared to numerical solutions of the Fokker-Planck equation), although the impact on the presently used coefficients is for the most part minimal. 
The thermal conductivities are calculated as
\begin{equation}
\kappa_{e,i} = \frac{20 \left( \frac{2}{\pi} \right)^{3/2} (kT)^{5/2} k \tau_{e,i} \Xi_{B,e,i} }{ (m^{0.5} e^4 Z~\text{ln}\Lambda_{ei,ii}) } ,
\end{equation}
\begin{equation}
\tau_e = \frac{ 0.43 \bar Z }{ 3.44 + \bar Z + 0.26 \log{\bar Z} } ,
\end{equation}
\begin{equation}
\tau_i = 0.162 ,
\label{tau}
\end{equation}
\begin{equation}
\label{eqn:eh}
\Xi_{B,e} = \frac{1} {3.2021} \left[ \frac{ 6.18 + 4.66 x_e }{ (1.93 +2.31 x_e + 5.35 x_e^2 + x_e^3 } \right] ,
\end{equation}
\begin{equation}
\Xi_{B,i} = \frac{1} {3.8823} \left[ \frac{2 x_i^2+2.64}{x_i^4 + 2.7 x_i^2 + 0.68} \right] ,
\end{equation}
in which $x_i$ is the target ion Hall parameter. The energy transfer terms are calculated as
\begin{equation}
P_{hc,i} = \kappa_i A_0 \frac{T_{T,i}-T_{L,0}}{\delta_1 R_T} ,
\end{equation}
\begin{equation}
P_{hc,e} = \kappa_e A_0 \frac{T_{T,e}-T_{L,0}}{\delta_1 R_T} ,
\end{equation}
in which $\delta_1$ is a scaling factor to set the scale length for calculating the gradient. In the current studies it is set always to 0.25, which is similar to the 0.2 used by Lindemuth and gives fair agreement with 1D HELIOS\cite{macfarlane06}  simulation results for a cooling-slab problem.

For depositing the heat conducted from the target into the liner, we assume it is deposited uniformly throughout the liner, such that
\begin{equation}
P_{hc,j} = \frac{P_{hc,e} + P_{hc,i}} {N}.
\end{equation}

As mentioned in Section \ref{sec:mag}, the coupling of Nernst effect and field advection in the target may lead to a more Bohm-like thermal conductivity, such that it decreases only as $x^{-1}$ rather than $x^{-2}$ in the high-field limit. As such, we also include the option of using a modified thermal conductivity that decreases as $x^{-1}$, wherein we use Eq.~$\left(\ref{eqn:eh}\right)$ for both electrons and ions with the $x^3$ term in the denominator removed.

\subsubsection{End Losses}

In a spherical target with an azimuthal field and axial current, the field strength would go to zero on axis. This would result in increased thermal conduction losses (as well as increased alpha particle escape) at the poles of the target. We allow for the possibility of enhanced end losses by introducing a user-variable loss area $A_{end}$ in which the heat loss is unimpeded by magnetization,
\begin{equation}
P_{end,i} = \kappa_{i0} A_{end} \frac{T_{T,i}-T_{L,0}}{\delta_1 R_T} ,
\end{equation}
\begin{equation}
P_{end,e} = \kappa_{e0} A_{end} \frac{T_{T,e}-T_{L,0}}{\delta_1 R_T} ,
\end{equation}
in which $\kappa_{i0}$ and $\kappa_{e0}$ are the unmagnetized thermalized conductivities. 

Introducing full unmagnetized losses raises the issue of nonlocal electron heat conduction, in which it is seen that calculations using the local gradient and thermal conductivity overestimate the heat flux in ICF targets by an order of magnitude or so. The heat reduction increases with the product of electron free path and perturbation wavenumber.\cite{epperlein1990kinetic} In the proposed PJMIF regime, the electron delocalization length scale $\lambda_e = T^2 / 4 \pi n e^4$ is increased in comparison to ICF targets due to the decreased density, but the wavenumber is decreased due to the increased gradient length scale. We model these effects by including a flux limiter for the heat conduction, such that heat transport is not allowed to exceed a fraction $f$ of the thermal free-streaming value,
\begin{equation}
P_{end,e}  \leq f \frac{3^{3/2}}{2} \frac{n_{T} {T_{T,e}}^{3/2}} {\sqrt{m_e}} .
\end{equation}

\subsubsection{Radiation}

The reduced target densities of MIF as compared to ICF mean that the effects of radiation pressure are typically negligible throughout the implosion. As the opacities of the MIF targets may remain relatively low, radiation is important mainly in its capacity to transport energy out of the target and into the liner. We thus take the radiation rate as a flux-limited conductive term in the diffusion limit,\cite{turner2001module} and similarly to the heat conduction term we deposit it uniformly across liner zones.
\begin{equation}
P_{rad} = D_{r} A_0 \frac{E_{r,T}-E_{r, \bar L}}{\delta_1 R_T} ,
\end{equation}
\begin{equation}
D_{r} = \frac{c}{3\chi} ,
\end{equation}
\begin{equation}
\chi \approx 10^2 \rho ,   
\end{equation}
\begin{equation}
P_{rad,j} = \frac{P_{rad}}{N} ,
\end{equation}
in which $E_{r} = e_{r} \rho$ is the radiation energy density.

\subsubsection{Equilibration}

We consider energy transfer within the target between ions and electrons and between electrons and the radiation field.  The coupling terms used are\cite{nrl-formulary, macfarlane06}
\begin{equation}
P_{ie} = \nu_{ie} m_T \left(e_{e,T} - e_{i,T}\right) ,
\end{equation}
\begin{equation}
\nu_{ie} = 1.8~\text{x}~10^{-19} \frac{\sqrt{m_i m_e} n \lambda_{ie}} {(m_i T_e  +  m_e T_i)^{1.5}} ,
\end{equation}
\begin{equation}
P_{er} =  \frac{8 \pi^5}{15}\frac{ \left(k_B T_{e,T}\right)^4 \sigma_T m_T}{c^2 h^3} - c \sigma_T E_{r,T} ,
\end{equation}
in which $\lambda_{ie}$ is the Coulomb logarithm and $\sigma_T$ is the target opacity, for which we use a Kramer's Law opacity,\cite{carroll2006introduction}
\begin{equation}
\sigma_T = 10^{-14}n_T T_{e,T}^{-3.5} .
\end{equation}

\subsubsection{Fusion reactions}

D-T fusion reaction yields are modeled using the functional form of Bosch and Hale\cite{bosch1992improved} as given by McBride and Slutz.\cite{mcbride15}

\subsubsection{Charged-particle deposition}

We assume instantaneous energy deposition of a fraction of the fusion-born 3.5-MeV alpha particles based on the magnetic field strength and stopping length in the target. The alpha-energy-deposition fraction is precomputed using a non-relativistic Monte-Carlo particle-tracking approach using the Lorentz force and dynamical friction slowing only, i.e., neglecting velocity-space diffusion. Monte-Carlo data and discussion are included in Appendix~\ref{app:alpha}. 

Fitting formulae for the alpha-deposition fraction are devised in the normalized variable notation of Basko,\cite{basko00} $\bar R = R_T / l_a$, where $l_a$ is the alpha collisional stopping length, and $b = R_T / r_{c,\alpha}$, in which $r_{c,\alpha}$ is the 3.5-MeV alpha cyclotron radius, for the case of the field generated by a uniform current along a central axis of the spherical target. For the field generated due to a uniform current along the central axis, the fit formula we find is
\begin{equation}
x = \frac{3}{2} \bar R \left[ 1 + \left(\frac{b}{3}\right)^{\frac{4}{3}} \right] ,
\label{eqn:alpha1}
\end{equation}
\begin{equation}
f_{\alpha,j} = \frac{x + x^2}{ 1 + \frac{13}{9} x + x^2 } .
\label{eqn:alpha2}
\end{equation}
To approximate the collisional alpha stopping length in the target plasma, we consider the slowing rate on the target electrons and ions,\cite{nrl-formulary}
\begin{equation}
\nu_{\text{fast}} = \num{1.7e-4}~\mu_\alpha^{0.5} \epsilon_\alpha^{-1.5} n_T Z_\alpha^2 \lambda_{ie},
\end{equation}
\begin{equation}
\nu_{\text{slow}} = \num{1.6e-9}~\mu_\alpha^{-1} T_T^{-1.5} n_T Z_\alpha^2 \lambda_{ie}.
\end{equation}
\begin{equation}
\label{eqn:stopping}
l_\alpha = v_{\alpha0} \left( \frac{1}{\nu_{\text{fast}}}  + \frac{1}{\nu_{\text{slow}}} \right).
\end{equation}
in which $v_{\alpha0} \approx~\num{1.3e9}$ cm/s is the initial speed of the 3.5 MeV alpha particle. Deposited energy is split between target electrons and ions as\cite{fraley1974thermonuclear}
\begin{equation}
\frac{f_{\alpha,i}}{f_\alpha} = \frac{T_{T,e} \text{[keV]}}{32 + T_{T,e} \text{[keV]}} ,
\end{equation}
\begin{equation}
\frac{f_{\alpha,e}}{f_\alpha} = \frac{32}{32 + T_{T,e} \text{[keV]}} .
\end{equation}
For alpha energy that escapes the target, we deposit it in the liner if the collisional slowing length in the liner is less than the liner thickness. We implement this as the following fractional deposition:
\begin{equation}
f_{\alpha,L} = \frac{1}{1+\left( \frac{l_{\alpha,L}}{R_L} \right)^2} .
\end{equation}

\subsection{Liner formation and convergence}
\label{sec:form_and_conv}

The first phase of liner formation and convergence is assumed to begin from discrete plasma jets arrayed around a spherical vacuum vessel. The machine size is dictated by the requirement to keep a manageable heat load on the first wall when operating at frequency $f_{\text{rep}}$ at a gain $G$. We restrict the average heat load $H$ to set the first wall radius
\begin{equation}
R_w = \sqrt{\frac{ \left ( 1+G \right ) E_{\text{tot}} f_{\text{rep}}} {4 \pi H}}.
\end{equation}
For the studies in this paper, values assumed are $f_{\text{rep}} =$ 1~Hz, $H = 2.5~\text{MW/m}^2$. It is assumed that the jets are injected at the first wall radius, $r_{ji} = R_w$. The merging radius for a given number of guns was calculated by Cassibry et al.\cite{cassibry13} who assumed that the jets expand at a speed of $2 C_s / (\gamma - 1)$. Taking the expansion speed of the jet as $C_s$ as suggested in experiments,\cite{hsu12pop} we use a similar form derived with this assumption
\begin{equation}
R_m = \left ( \frac{ r_{j0} (M_j + 1) + R_{ji} } { 1 + \frac{2}{\sqrt{N_j}} \left( M_j+1 \right)} \right),
\end{equation}
in which $N_j$ is the number of jets, $r_{j0}$ is the initial radius of the jet plasma column, and $M_j$ is the jet mach number.

At the merging radius, oblique shocks may occur as the jets interact, which may lead to heating and/or non-uniformities in the liner.\cite{merritt13,merritt14} In 1D, we neglect the issue of non-uniformities but consider heating. Shocks may heat the ions, degrade the liner Mach number, and cause increased spreading of the liner material for the remainder of the travel to the target. In the case of a high-$Z$ liner, however, ionization, ion equilibration with electrons, and radiative cooling may combine to keep the liner temperature below a few eV.

Combining pre- and post-merge travel of the liner, we can approximate the radial expansion of the liner into vacuum as:
\begin{equation}
\label{eqn:deltaL}
\Delta L =  2 \bar C_{s,bm}\frac{R_{ji} - R_{m}}{u_0} + 2 \bar C_{s,am}\frac{R_{m} - R_{T}}{u_0},
\end{equation}
where $\bar C_{s,bm}$ and $\bar C_{s,bm}$ represent the average sound speeds before and after merge and the factors of 2 reflect the fact that the jet may expand inward towards the target as well as outwards towards the chamber wall. 
A more-accurate estimate of the radial liner expansion would need to account for time dependence of the 
liner sound speed and the 3D dynamics of the merging jets.

\section{Results and discussion}
\label{sec:results}

\subsection{Verification}

\begin{figure}[!htb]
\begin{center}
\includegraphics[width=3.34in]{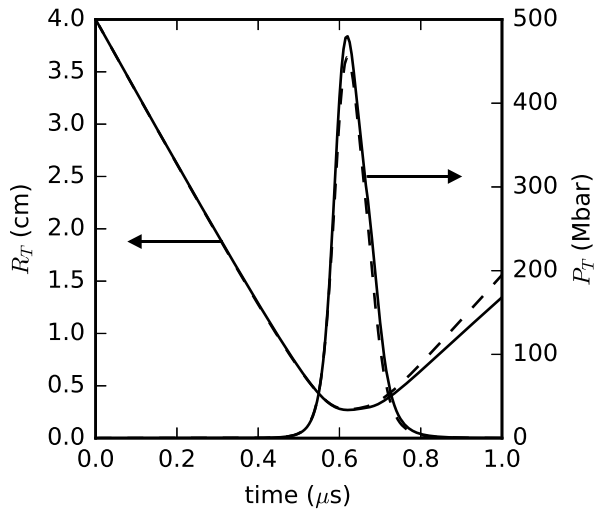}
\caption{Comparison between present model (dashed lines) and HELIOS (solid lines) for a nominal PJMIF target compression, case1 of Table~\ref{tab:init_conditions}.}
\label{fig:comparison1}
\end{center}
\end{figure}

Verification cases are run to confirm that the model is working correctly. Results are primarily compared to the well-benchmarked 1D hydrocode HELIOS from Prism Computational Sciences.\cite{macfarlane06} Figure~\ref{fig:comparison1} shows a comparison between model and HELIOS results for stagnation pressure and target convergence for a nominal PJMIF target compression. Additional verifications are shown in Appendix~\ref{app:verification}.

The numerical integration is carried out using a Heun predictor-corrector iteration with one forward Euler prediction step and one trapezoidal rule correction step. Energy conservation in the model is typically maintained within 1\%. Typically,\ 40 zones are used in the liner.

\begin{table}
\caption{\label{tab:init_conditions}Initial conditions for specific PJMIF target-compression cases.}
\begin{ruledtabular}
\begin{tabular}{ccccc}
&case1&case2\\ 
 \hline
 $E_0$ (MJ)&18.8&27.6 \\
 $\rho_L$ (g/$\text{cm}^{-3}$)&0.03&0.09 \\
 $T_{L}$ (eV)&1.5&1.5 \\
 $L_{L}$ (cm)&1.0&1.0 \\
 $u_{L}$ (cm/$\mu$s)&7&5.5 \\
 $\mu_{L}$ (amu)&131.2 (Xe)&131.2 (Xe) \\
 $\rho_{T}$ (g/$\text{cm}^{-3}$)&$10^{-5}$&$6\times10^{-5}$ \\
 $T_{T,i}$ (eV)&100&100 \\
 $T_{T,e}$ (eV)&100&100 \\
 $T_{T,rad}$ (eV)&10&10 \\
 $R_{T}$ (cm)&4.0&3.5 \\
 $\bar B_{T} (T)$&4.39&10.8 \\
 $\beta_{T}$&10&10 \\
 \end{tabular}
\end{ruledtabular}
\end{table}

\subsection{Target-compression phase}

The target-compression model as formulated uses twelve input parameters to define the liner and target: $\rho_L,~T_L,~L_L,~u_L,~\mu_L,~Z,~\rho_T,~T_{T,e},~T_{T,i},~T_{T,r},~R_T,$ and $B_T$ or $\beta_T$. Rather than extensively scanning the large parameter space here, we focus on a few nominal cases of interest for PJMIF and consider variations of the model physics settings and initial conditions.

Table~\ref{tab:physics_variations} shows the impact of different physics assumptions on the results of a PJMIF target compression. The case (case2 from Table~\ref{tab:init_conditions}) is selected to rely on magnetized alpha deposition to achieve gain greater than 20 with a minimum of liner dynamic pressure. For these cases the flux limiter for end losses is set to $f$ = 0.06 and $\delta_1$ is set to 0.25. As seen in Table~\ref{tab:physics_variations}, case 2a shows that a gain greater than 20 is achievable with magnetized alpha deposition. Much of the stagnation pressure derives from alpha deposition and burn, as the liner dynamic pressure only needs to reach 450 Mbar before the alpha deposition initiates a rapid pressure rise. Case 2b) underscores this dependence on alpha heating. Results of assuming Bohm-like transport rather than classical are shown in case 2c), which drops the gain to near unity. Results of increased end losses in case2d) show a similar effect, dropping gain to 2.4. Case 2a) is thus positioned near somewhat of an 'ignition cliff' in parameter space, where any increase in losses leads to a significant reduction in gain.

\begin{table*}
\caption{\label{tab:physics_variations}Model results for varied physics settings case2 of Table~\ref{tab:init_conditions}:}
\begin{ruledtabular}
\begin{tabular}{ccccc}
&2a)&2b)&2c)&2d)\\
 heat conduction&Braginskii&Braginskii&Bohm-like&Braginskii\\ 
 alpha deposition&yes&no&yes&yes\\ 
 end loss $A_{end} / A_0$ & 0.005 & 0.005  & 0.005 & 0.010 \\
 $E_0$ (MJ)&27.7&27.7&27.7&27.7 \\
 \hline
 $\text{CR}_{\text{max}}$     &   12.5    &  28.0    &  23.7   &  14.5  \\
 $T_{\text{i,T,max}}$ (keV)   &   20.8    &  3.99    &  3.94   &  4.98  \\
 $P_{\text{T,max}}$ (Mbar)   &  1330    &  676    &  622    &  725   \\
 yield ($10^{19}$ n)              &  21.7     &  0.567 &  14.4   &  2.36  \\
 gain                                     &  22.0     &  0.577  &  0.84  &   2.41 \\
 \end{tabular}
\end{ruledtabular}
\end{table*}

Considering further perturbations to the case of Table~\ref{tab:physics_variations}, Table~\ref{tab:ic_variations} shows results of changes in the liner and target initial conditions while keeping the overall energy constant.  It is seen in case 2e) that increased thickness of the liner results in a substantial drop in achieved stagnation pressure and gain, dropping gain below unity even if one compensates somewhat by adjusting the target mass (case 2f). Changes to the liner temperature and molecular weight are shown in cases 2g) and 2h), which also significantly reduce gain to 0.87 and 3.3 respectively.

\begin{table*}
\caption{\label{tab:ic_variations}Variations of initial conditions of case2 of Table~\ref{tab:init_conditions}:}
\begin{ruledtabular}
\begin{tabular}{ccccc}
&2e)&2f)&2g)&2h)\\
 &2x liner thickness&2x liner thickness,&4x liner temperature&$\mu_L = 39.948\ \text{(Ar)}$\\ 
 & &0.5x target mass& & \\
 \hline
 heat conduction&Braginskii&Braginskii&Braginskii&Braginskii\\ 
 alpha deposition&yes&yes&yes&yes\\ 
 end loss $A_{end} / A_0$ & 0.005 & 0.005  & 0.005  & 0.005 \\
 $E_0$ (MJ)                       & 27.7   &27.6     & 27.9    & 27.8 \\
 \hline
 $\text{CR}_{\text{max}}$  &17.9     &20.7     & 12.0     &12.3 \\
 $T_{\text{i,T,max}}$ (eV)  &2.88     &4.55     & 4.31     & 6.32 \\
 $P_{\text{T,max}}$ (Mbar)&107     &158       & 330       & 542 \\
 yield ($10^{19}$ n)           & 0.098   &0.277     & 0.859        &3.29 \\
 gain                                   &0.100   &0.276     & 0.867        &3.33 \\
 \end{tabular}
\end{ruledtabular}
\end{table*}

\begin{figure}[!htb]
\begin{center}
\includegraphics[width=3.34in]{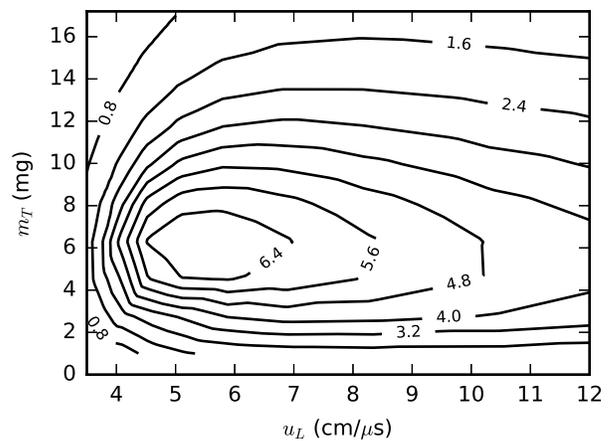}
\caption{Contours of fusion gain for target compressions assuming classical transport, alpha deposition on, for $E_0$ = 20~MJ, $R_L$ = 1.0 cm, $R_T$~= 4.0 cm, $\beta_T$ = 10, $x_{T,e}$ = 10,  $\mu_{L}$ = 131.2, $T_{L}$ = 1.5 eV. Here, gain is defined as the fusion energy divided by the liner and target input energies.}
\label{fig:gaincontours_brag}
\end{center}
\end{figure}

\begin{figure}[!htb]
\begin{center}
\includegraphics[width=3.34in]{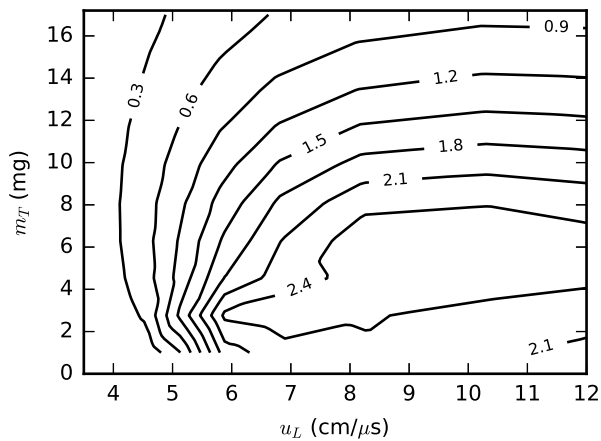}
\caption{Contours of fusion gain for target compressions assuming Bohm-like transport, alpha deposition on, $E_0$ = 20~MJ, $R_L$ = 1.0 cm, $R_T$~= 4.0 cm, $\beta_T$ = 10, $x_{T,e}$ = 10,  $\mu_{L}$ = 131.2, $T_{L}$ = 1.5 eV.}
\label{fig:gaincontours_bohm}
\end{center}
\end{figure}

Figure~\ref{fig:gaincontours_brag} shows a scan of implosion speeds and target densities for a given constant input energy of 20~MJ and fixed liner and target thicknesses. Figure~\ref{fig:gaincontours_bohm} repeats the scan, but with the assumption of Bohm-like thermal conduction losses as described in Sec.~\ref{sec:therm}. In agreement with intuition, the increased losses in the latter case cause a decrease in gain and increase the optimal implosion speed. In the Bohm-like transport case, maximal gain is attained at around 9 or 10 cm/$\mu$s implosion speed, compared to about 5.5 cm/$\mu$s in the classical-transport cases.

\subsection{Liner formation and convergence phase}

From the investigation of the target-compression phase, it is found that it is desirable to achieve a concentration of the liner mass and kinetic energy into a liner thickness of order 1 cm. The requirements for this can be evaluated from Eq.~(\ref{eqn:deltaL}) and the system constraints outlined in Sec.~\ref{sec:form_and_conv}.

\begin{figure}[!htb]
\begin{center}
\includegraphics[width=3.34in]{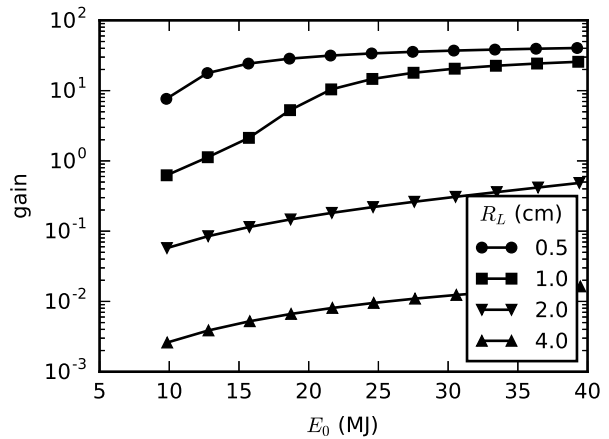}
\caption{Gain vs input energy for differing liner thicknesses, with classical transport, alpha deposition on, $m_L$ = 0.75 g/MJ (roughly $u_L$ = 5 cm/$\mu$s,) $m_T$ = 0.3, 0.25, 0.20, 0.15 mg/MJ respectively for $R_L$ = 0.5, 1.0, 2.0, 4.0 cm, $R_T$~= 3.5 cm, $\beta_T$ = 10, $x_{T,e}$ = 10,  $\mu_{L}$ = 131.2, $T_{L}$ = 1.5 eV.}
\label{fig:RL_gain}
\end{center}
\end{figure}

\begin{figure}[!htb]
\begin{center}
\includegraphics[width=3.34in]{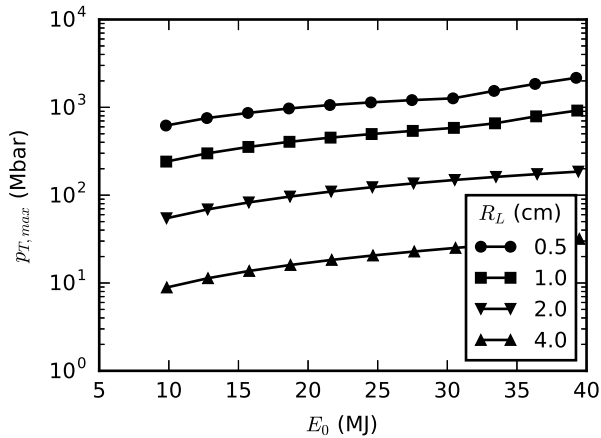}
\caption{Max no-burn target pressure vs input energy for differing liner thicknesses, with classical transport, alpha deposition, $m_L$ = 0.75 g/MJ (roughly $u_L$ = 5 cm/$\mu$s,) $m_T$ = 0.3, 0.25, 0.20, 0.15 mg/MJ respectively for $R_L$ = 0.5, 1.0, 2.0, 4.0 cm, $R_T$~= 3.5 cm, $\beta_T$ = 10, $x_{T,e}$ = 10,  $\mu_{L}$ = 131.2, $T_{L}$ = 1.5 eV.}
\label{fig:RL_pmax}
\end{center}
\end{figure}

For the parameters of case2 of Table~\ref{tab:init_conditions}, at the energy of 31.3 MJ, the gain of 23.5, and the assumed 1-Hz repetition rate and a 2.5-MW/$\text{m}^2$ heat load (rather conservative for a liquid first wall),
the required first-wall radius is 5 m. Assuming a xenon liner with average temperature 1 eV during both pre- and post-merge transit, and a liner velocity of 70 km/s corresponding to an average liner Mach number of 63, the estimated radial expansion of the liner is 15 cm by the time of the target encounter.  Clearly, such a liner would achieve much lower stagnation pressures than the liners of 1-2 cm thickness that have shown promising performance in the target compression analysis. It is evident that if it is desired to reach gain $>$ 10 without an afterburner at these energies, PJMIF needs to employ techniques to lessen the radial expansion of the liner during the transit of the standoff distance. Figures~\ref{fig:RL_gain} and \ref{fig:RL_pmax} further elucidate this effect, with Fig.~\ref{fig:RL_pmax} showing clearly the reduction in achievable stagnation pressures associated with the liner spreading.


\section{Conclusions}

In this paper, we have formulated a semi-analytic, reduced 1D physics model for PJMIF and applied it to study 
plasma-liner implosions and target fusion gains without a cold fuel layer or ``afterburner.'' The results indicate 
that target gains as high 30 might be possible if plasma liners as thin as 0.5 cm can be formed, and yielded 
additional physical insights into the concept. Gain is attainable for both classical and Bohm-like heat transport,
with or without alpha energy deposition, which highlights the potential advantage of the high implosion speeds
of a plasma liner in mitigating transport-based losses.

Our consideration of the PJMIF liner-formation and convergence phases support (i) the original
viewpoint of \textcite{thio99} that the use of an afterburner to amplify the target fusion gain (with the target acting as a Òhot 
spotÓ) is an essential part of the PJMIF concept, unless plasma liners with thickness of 1 cm or less can be formed,
and (ii) the desirability of forming a plasma liner as thin as a few centimeters in order to produce sufficient target fusion 
yield for achieving significant burn in the afterburner. 

A useful extension of the model would be to develop an approach for non-local and non-instantaneous alpha deposition, 
in order to study propagating burn waves in the target and afterburner.
More broadly, fruitful future 1D PJMIF work may seek solutions 
to the problem of jet/liner expansion during the transit of the standoff distance, employ this model or a variant to scan 
large regions of parameter space for optimal configurations, investigate additional fuel layers for higher gain, or study 
suitable magnetized-target formation.

\appendix
\section{Alpha-Deposition}
\label{app:alpha}
\begin{figure}[!htb]
\begin{center}
\includegraphics[width=3.34in]{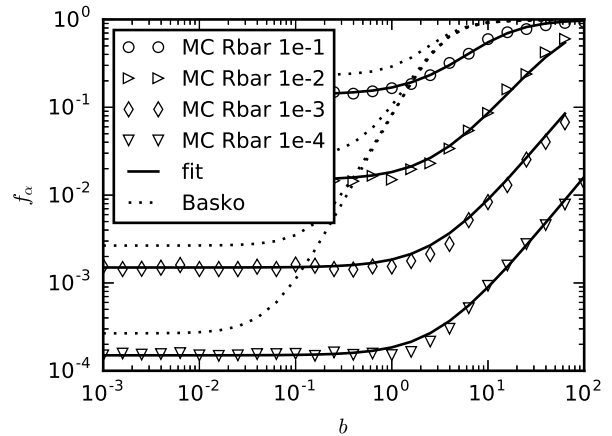}
\caption{Comparison between alpha deposition fitting Eqs.~(\ref{eqn:alpha1} and \ref{eqn:alpha2}) for axial-current spherical geometry, our dynamical-friction-slowing Monte Carlo results, and the fitting formula of Basko et al.\ for axially magnetized cylindrical geometry.}
\label{fig:alphas}
\end{center}
\end{figure}

Equations (\ref{eqn:alpha1}) and (\ref{eqn:alpha2}) are generated from Monte Carlo calculations employing dynamical-friction slowing of alpha particles and the Lorentz force. We include the plot results and fit Eqs.~(\ref{eqn:alpha1} and \ref{eqn:alpha2}) in Fig.~\ref{fig:alphas}. It is interesting to compare the present results to the Basko et al. formula,\cite{basko00} so it is included Figure~\ref{fig:alphas}.

It is perhaps surprising that in these results the dependence on collisional stopping length in $\bar R$ persists in the limit of strong magnetization. From inspection of particle histories, we see that the dependence arises due to a loss cone in the central axis of the spherical target with axial magnetizing current, where the field crosses zero. Under the assumptions of dynamical-friction slowing only, alphas that cross close enough to the axis can enter into a switchback-type of trajectory, and ultimately escape the target at one of the poles. In this way, even a very strong magnetic field can be escaped by an increasing fraction of particles if the stopping distance is long enough. This is ultimately the reason that the dependence on $\bar R$ remains in the fitting formula for strongly magnetized cases. 

By contrast, in the situation considered by Basko et al.\cite{basko00} the cylindrical target is infinite in length and there are no poles through which to escape, and the field is all in the same direction so the switchbacks cannot occur.

\section{Benchmarking against HELIOS}
\label{app:verification}
We compared results from the present model against HELIOS for several test problems. While HELIOS does not contain all of the features explored in the present model, it is a reliable hydrocode in 1D spherical geometry with options to treat radiation, thermal conduction, ionization / EOS, and charged particle energy deposition, and thus can be used to verify that the numerical aspects of the model are functioning correctly. Figures~\ref{fig:iz} and \ref{fig:comparison1} have shown the performance of the hydrodynamic and ionization routines, so this section will address other physics areas individually before examining full runs with all physics turned on.

\subsection{Ion-electron equilibration}
\begin{figure}[!htb]
\begin{center}
\includegraphics[width=3.34in]{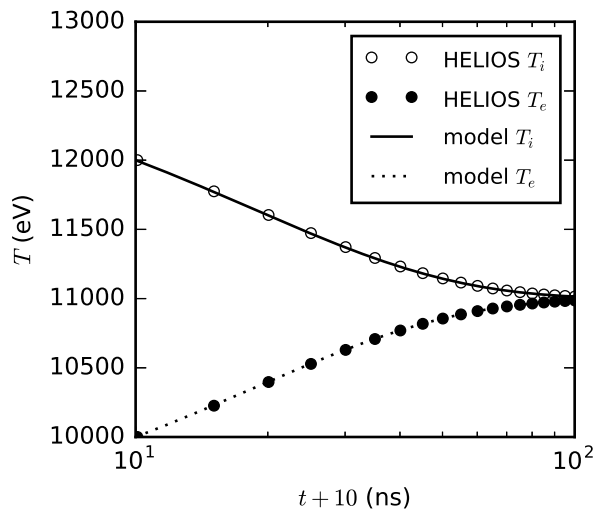}
\caption{HELIOS / model comparison of ion-electron equilibration rates of a 1-cm radius DT sphere, $\rho = 0.01~\text{g/cm}^3$.}
\label{fig:comparison3}
\end{center}
\end{figure}

To check the implementation of separate ion and electron temperatures in the target, a DT sphere of 1-cm radius is initialized at different ion and electron temperatures with hydrodynamics disabled and allowed to thermally equilibrate. As shown in Fig.~\ref{fig:comparison3}, HELIOS and the present model agree closely.

\subsection{Heat conduction}
\begin{figure}[!htb]
\begin{center}
\includegraphics[width=3.34in]{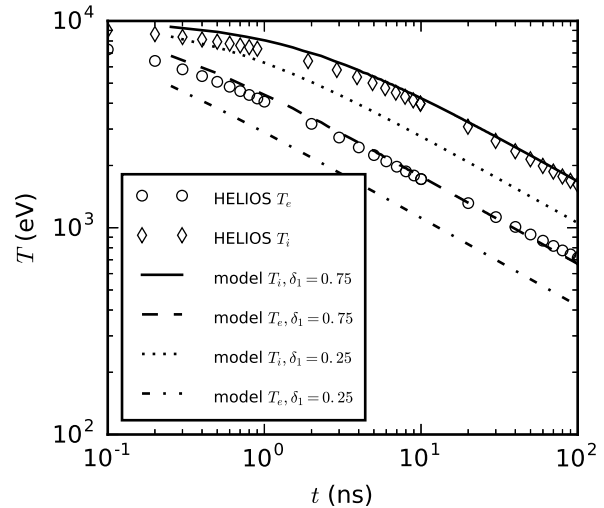}
\caption{HELIOS / model comparison of ion and electron thermal conduction loss rates of a 1 cm radius DT sphere, $\rho = 0.01~\text{g/cm}^3$, $\tau_i = 1$.}
\label{fig:comparison4}
\end{center}
\end{figure}

HELIOS includes a Spitzer heat conduction feature in 1D spherical geometry, which can be used to check the thermal conduction losses and help determine appropriate settings for the gradient-scale-length-factor $\delta_1$ in the model. With hydrodynamics off, a DT sphere of 1-cm radius is initialized next to a cold liner and allowed to lose energy via conduction. Electron-ion equilibration is also turned off to isolate each species. As shown in Fig.~\ref{fig:comparison4}, HELIOS and the present model behave quite similarly, confirming that the two-temperature treatment and the heat-conduction treatment are implemented correctly. It is shown that different values of $\delta_1$ give the best result for different timescales of interest.
We note that HELIOS uses $\tau_i$ = 1 for the equivalent of Eq.~(\ref{tau}) rather than $\tau_i$ = 0.162, which we find to be the classical / Braginskii value. For these comparisons to HELIOS, we use $\tau_i$ = 1 in the model. For results reported elsewhere in this paper, we use Eq.~(\ref{tau}), $\tau_i$ = 0.162.

\subsection{Alpha deposition}
\begin{figure}[!htb]
\begin{center}
\includegraphics[width=3.34in]{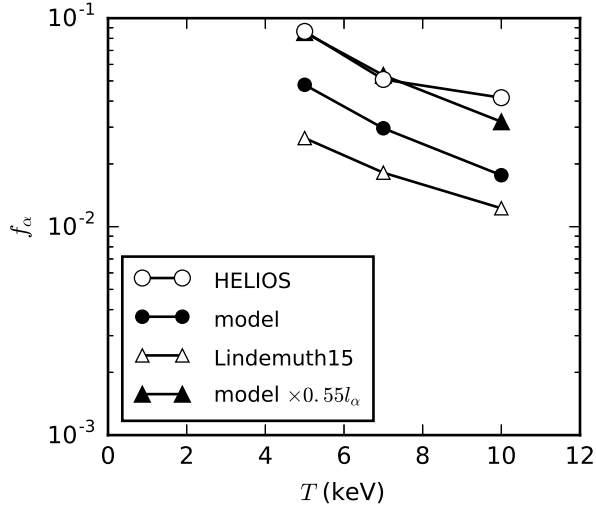}
\caption{HELIOS / model comparison of alpha deposition fractions.}
\label{fig:comparison_alpha}
\end{center}
\end{figure}

To check the transition from the non-dimensionalized Monte Carlo model to real conditions, we compare model results of a 1-cm radius DT sphere to HELIOS's charged particle stopping power model as well as the analytical formula for collisional alpha deposition fraction included in Lindemuth 2015.\cite{lindemuth15} As shown in Fig.~\ref{fig:comparison_alpha}, the models are in coarse agreement. When it is desired to benchmark against HELIOS in cases with alpha deposition, we find that adjusting the model stopping length by a factor of 0.55 gives good agreement. 

\subsection{Target compressions}
\begin{figure}[!htb]
\begin{center}
\includegraphics[width=3.34in]{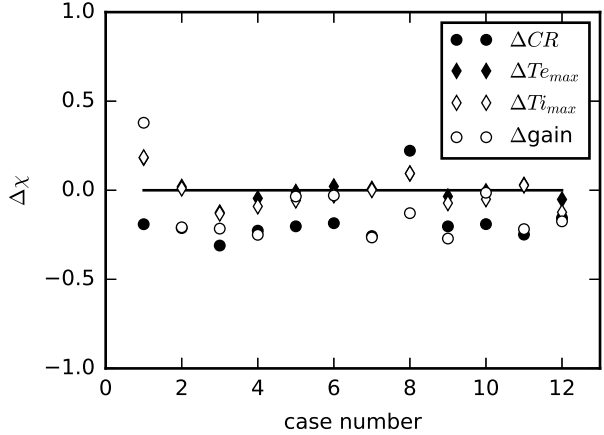}
\caption{HELIOS / model comparison of results for gain, convergence ratio, and target ion / electron temperatures for cases of Table~\ref{tab:helios_bench}. Differences in results are calculated as $\Delta\chi = (\chi_{model} - \chi_{HEL}) / \chi_{HEL}$. Mass-weighted average target temperatures are used for the HELIOS runs.}
\label{fig:comparison_helios}
\end{center}
\end{figure}

To check that the diverse physics in the model are not causing problems when combined together, we compare some randomly specified HELIOS runs with all relevant physics enabled to model runs with matching settings. Liner thickness is held fixed at 1 cm and initial target radius = 4 cm. HELIOS Physics options are set to use radiation diffusion, tabular EOS, 0.1\% Spitzer heat conduction, and 20\% of alpha energy deposited directly into the ion fluid with the rest escaping. Model features are set the same as well as using $\tau_i$ = 1, $\delta_1 = 0.25$, and $0.55 l_\alpha$.  The results are listed in Table~\ref{tab:helios_bench} and the differences are plotted in Fig.~\ref{fig:comparison_helios}. It is seen that the model and HELIOS reach similar answers across the board, verifying that the model implementation is behaving as anticipated.

\begin{table*}
\caption{\label{tab:helios_bench}Comparison between model and HELIOS over several cases.}
\begin{ruledtabular}
\begin{tabular}{c|ccccc|cccc|cccc}
\multicolumn{6}{c|}{Initial Conditions}&\multicolumn{4}{c|}{Model Results}&\multicolumn{4}{c}{HELIOS Results}\\
&$E_0$&$\rho_L$&$u_{L}$&$\rho_{T}$ &$T_{T}$                            &gain&$CR_{\text{max}}$&$T_{i,\text{max}}$&$T_{e,\text{max}}$&gain&$CR_{\text{max}}$&$T_{i,\text{max}}$&$T_{e,\text{max}}$\\ 
&(MJ)&(g/$\text{cm}^{-3})$&(cm/$\mu$s)&(g/$\text{cm}^{-3}$)&(eV)&  ( )   &  ( ) &(keV)&(keV)& ( )  &  ( ) &(keV)&(keV)\\ 
 \hline
1 & 29.9 & 0.1683 & 4.24 & 4.74e-05 & 115.1 & 23.28 & 12.24 & 14.46 & 11.82 & 16.88 & 15.12 & 12.21 & 10.00 \\
2 & 27.7 & 0.1253 & 3.66 & 4.12e-05 & 132.8 & 3.10 & 10.35 & 6.07 & 5.83 & 3.92 & 13.14 & 6.01 & 5.72 \\
3 & 27.0 & 0.0595 & 5.22 & 5.65e-05 & 130.2 & 2.19 & 8.83 & 5.59 & 5.41 & 2.79 & 12.80 & 6.43 & 6.18 \\
4 & 23.5 & 0.0682 & 6.22 & 5.85e-05 & 142.0 & 9.23 & 9.89 & 9.98 & 9.02 & 12.30 & 12.80 & 10.97 & 9.44 \\
5 & 23.7 & 0.0710 & 6.39 & 3.84e-05 & 128.7 & 18.48 & 11.95 & 17.79 & 13.35 & 19.15 & 14.99 & 18.87 & 13.48 \\
6 & 33.9 & 0.1416 & 4.14 & 4.01e-05 & 117.8 & 11.43 & 12.06 & 9.95 & 8.98 & 11.78 & 14.79 & 10.24 & 8.79 \\
7 & 23.0 & 0.0688 & 6.22 & 8.81e-05 & 152.9 & 3.77 & 8.50 & 6.33 & 6.09 & 5.13 & 11.44 & 6.33 & 6.04 \\
8 & 35.3 & 0.2186 & 3.69 & 7.50e-05 & 128.3 & 1.51 & 24.53 & 4.34 & 4.25 & 1.73 & 20.07 & 3.97 & 3.88 \\
9 & 20.3 & 0.0544 & 5.37 & 2.79e-05 & 106.9 & 5.29 & 12.07 & 8.80 & 8.06 & 7.25 & 15.14 & 9.49 & 8.35 \\
10 & 37.6 & 0.0676 & 6.08 & 2.68e-05 & 114.4 & 15.56 & 13.57 & 17.24 & 12.96 & 15.77 & 16.76 & 18.16 & 13.04 \\
11 & 25.9 & 0.0612 & 5.15 & 5.02e-05 & 115.3 & 2.52 & 9.72 & 5.79 & 5.59 & 3.22 & 12.94 & 5.63 & 5.42 \\
12 & 26.4 & 0.0805 & 4.50 & 1.74e-05 & 101.8 & 6.94 & 15.19 & 10.51 & 9.21 & 8.40 & 17.99 & 11.99 & 9.72 \\
 \end{tabular}
\end{ruledtabular}
\end{table*}

\begin{acknowledgments}
We acknowledge Y. C. F. Thio, I. Golovkin, I. Lindemuth, R. Kirkpatrick, J. Cassibry, R. Samulyak, P. Stoltz, and K. Beckwith for helpful discussions.  This work was supported by the Advanced Research Projects Agency--Energy (ARPA-E) under Department of Energy contract no.\ DE-AC52-06NA25396.  
\end{acknowledgments}


\begin{thebibliography}{47}%
\makeatletter
\providecommand \@ifxundefined [1]{%
 \@ifx{#1\undefined}
}%
\providecommand \@ifnum [1]{%
 \ifnum #1\expandafter \@firstoftwo
 \else \expandafter \@secondoftwo
 \fi
}%
\providecommand \@ifx [1]{%
 \ifx #1\expandafter \@firstoftwo
 \else \expandafter \@secondoftwo
 \fi
}%
\providecommand \natexlab [1]{#1}%
\providecommand \enquote  [1]{``#1''}%
\providecommand \bibnamefont  [1]{#1}%
\providecommand \bibfnamefont [1]{#1}%
\providecommand \citenamefont [1]{#1}%
\providecommand \href@noop [0]{\@secondoftwo}%
\providecommand \href [0]{\begingroup \@sanitize@url \@href}%
\providecommand \@href[1]{\@@startlink{#1}\@@href}%
\providecommand \@@href[1]{\endgroup#1\@@endlink}%
\providecommand \@sanitize@url [0]{\catcode `\\12\catcode `\$12\catcode
  `\&12\catcode `\#12\catcode `\^12\catcode `\_12\catcode `\%12\relax}%
\providecommand \@@startlink[1]{}%
\providecommand \@@endlink[0]{}%
\providecommand \url  [0]{\begingroup\@sanitize@url \@url }%
\providecommand \@url [1]{\endgroup\@href {#1}{\urlprefix }}%
\providecommand \urlprefix  [0]{URL }%
\providecommand \Eprint [0]{\href }%
\providecommand \doibase [0]{http://dx.doi.org/}%
\providecommand \selectlanguage [0]{\@gobble}%
\providecommand \bibinfo  [0]{\@secondoftwo}%
\providecommand \bibfield  [0]{\@secondoftwo}%
\providecommand \translation [1]{[#1]}%
\providecommand \BibitemOpen [0]{}%
\providecommand \bibitemStop [0]{}%
\providecommand \bibitemNoStop [0]{.\EOS\space}%
\providecommand \EOS [0]{\spacefactor3000\relax}%
\providecommand \BibitemShut  [1]{\csname bibitem#1\endcsname}%
\let\auto@bib@innerbib\@empty
\bibitem [{\citenamefont {Lindemuth}\ and\ \citenamefont
  {Kirkpatrick}(1983)}]{lindemuth83}%
  \BibitemOpen
  \bibfield  {author} {\bibinfo {author} {\bibfnamefont {I.~R.}\ \bibnamefont
  {Lindemuth}}\ and\ \bibinfo {author} {\bibfnamefont {R.~C.}\ \bibnamefont
  {Kirkpatrick}},\ }\href@noop {} {\bibfield  {journal} {\bibinfo  {journal}
  {Nucl.\ Fusion}\ }\textbf {\bibinfo {volume} {23}},\ \bibinfo {pages} {263}
  (\bibinfo {year} {1983})}\BibitemShut {NoStop}%
\bibitem [{\citenamefont {Kirkpatrick}, \citenamefont {Lindemuth},\ and\
  \citenamefont {Ward}(1995)}]{kirkpatrick95}%
  \BibitemOpen
  \bibfield  {author} {\bibinfo {author} {\bibfnamefont {R.~C.}\ \bibnamefont
  {Kirkpatrick}}, \bibinfo {author} {\bibfnamefont {I.~R.}\ \bibnamefont
  {Lindemuth}}, \ and\ \bibinfo {author} {\bibfnamefont {M.~S.}\ \bibnamefont
  {Ward}},\ }\href@noop {} {\bibfield  {journal} {\bibinfo  {journal} {Fusion
  Tech.}\ }\textbf {\bibinfo {volume} {27}},\ \bibinfo {pages} {201} (\bibinfo
  {year} {1995})}\BibitemShut {NoStop}%
\bibitem [{\citenamefont {Thio}(2008)}]{thio08}%
  \BibitemOpen
  \bibfield  {author} {\bibinfo {author} {\bibfnamefont {Y.~C.~F.}\
  \bibnamefont {Thio}},\ }\href@noop {} {\bibfield  {journal} {\bibinfo
  {journal} {J. Phys.\ Conf.\ Ser.}\ }\textbf {\bibinfo {volume} {112}},\
  \bibinfo {pages} {042084} (\bibinfo {year} {2008})}\BibitemShut {NoStop}%
\bibitem [{\citenamefont {Lindemuth}\ and\ \citenamefont
  {Siemon}(2009)}]{lindemuth09}%
  \BibitemOpen
  \bibfield  {author} {\bibinfo {author} {\bibfnamefont {I.~R.}\ \bibnamefont
  {Lindemuth}}\ and\ \bibinfo {author} {\bibfnamefont {R.~E.}\ \bibnamefont
  {Siemon}},\ }\href@noop {} {\bibfield  {journal} {\bibinfo  {journal} {Amer.\
  J.\ Phys.}\ }\textbf {\bibinfo {volume} {77}},\ \bibinfo {pages} {407}
  (\bibinfo {year} {2009})}\BibitemShut {NoStop}%
\bibitem [{\citenamefont {Lindemuth}(2015)}]{lindemuth15}%
  \BibitemOpen
  \bibfield  {author} {\bibinfo {author} {\bibfnamefont {I.~R.}\ \bibnamefont
  {Lindemuth}},\ }\href@noop {} {\bibfield  {journal} {\bibinfo  {journal}
  {Phys.\ Plasmas}\ }\textbf {\bibinfo {volume} {22}},\ \bibinfo {eid} {122712}
  (\bibinfo {year} {2015})}\BibitemShut {NoStop}%
\bibitem [{\citenamefont {Basko}, \citenamefont {Kemp},\ and\ \citenamefont
  {\mbox{Meyer-ter-Vehn}}(2000)}]{basko00}%
  \BibitemOpen
  \bibfield  {author} {\bibinfo {author} {\bibfnamefont {M.~M.}\ \bibnamefont
  {Basko}}, \bibinfo {author} {\bibfnamefont {A.~J.}\ \bibnamefont {Kemp}}, \
  and\ \bibinfo {author} {\bibfnamefont {J.}~\bibnamefont
  {\mbox{Meyer-ter-Vehn}}},\ }\href@noop {} {\bibfield  {journal} {\bibinfo
  {journal} {Nucl.\ Fusion}\ }\textbf {\bibinfo {volume} {40}},\ \bibinfo
  {pages} {59} (\bibinfo {year} {2000})}\BibitemShut {NoStop}%
\bibitem [{\citenamefont {Slutz}\ \emph {et~al.}(2010)\citenamefont {Slutz},
  \citenamefont {Herrmann}, \citenamefont {Vesey}, \citenamefont {Sefkow},
  \citenamefont {Sinars}, \citenamefont {Rovang}, \citenamefont {Peterson},\
  and\ \citenamefont {Cuneo}}]{slutz10}%
  \BibitemOpen
  \bibfield  {author} {\bibinfo {author} {\bibfnamefont {S.~A.}\ \bibnamefont
  {Slutz}}, \bibinfo {author} {\bibfnamefont {M.}~\bibnamefont {Herrmann}},
  \bibinfo {author} {\bibfnamefont {R.}~\bibnamefont {Vesey}}, \bibinfo
  {author} {\bibfnamefont {A.}~\bibnamefont {Sefkow}}, \bibinfo {author}
  {\bibfnamefont {D.}~\bibnamefont {Sinars}}, \bibinfo {author} {\bibfnamefont
  {D.}~\bibnamefont {Rovang}}, \bibinfo {author} {\bibfnamefont
  {K.}~\bibnamefont {Peterson}}, \ and\ \bibinfo {author} {\bibfnamefont
  {M.}~\bibnamefont {Cuneo}},\ }\href@noop {} {\bibfield  {journal} {\bibinfo
  {journal} {Phys.\ Plasmas}\ }\textbf {\bibinfo {volume} {17}},\ \bibinfo
  {pages} {056303} (\bibinfo {year} {2010})}\BibitemShut {NoStop}%
\bibitem [{\citenamefont {Gomez}\ \emph {et~al.}(2014)\citenamefont {Gomez},
  \citenamefont {Slutz}, \citenamefont {Sefkow}, \citenamefont {Sinars},
  \citenamefont {Hahn}, \citenamefont {Hansen}, \citenamefont {Harding},
  \citenamefont {Knapp}, \citenamefont {Schmit}, \citenamefont {Jennings},
  \citenamefont {Awe}, \citenamefont {Geissel}, \citenamefont {Rovang},
  \citenamefont {Chandler}, \citenamefont {Cooper}, \citenamefont {Cuneo},
  \citenamefont {Harvey-Thompson}, \citenamefont {Herrmann}, \citenamefont
  {Hess}, \citenamefont {Johns}, \citenamefont {Lamppa}, \citenamefont
  {Martin}, \citenamefont {McBride}, \citenamefont {Peterson}, \citenamefont
  {Porter}, \citenamefont {Robertson}, \citenamefont {Rochau}, \citenamefont
  {Ruiz}, \citenamefont {Savage}, \citenamefont {Smith}, \citenamefont
  {Stygar},\ and\ \citenamefont {Vesey}}]{gomez14}%
  \BibitemOpen
  \bibfield  {author} {\bibinfo {author} {\bibfnamefont {M.~R.}\ \bibnamefont
  {Gomez}}, \bibinfo {author} {\bibfnamefont {S.~A.}\ \bibnamefont {Slutz}},
  \bibinfo {author} {\bibfnamefont {A.~B.}\ \bibnamefont {Sefkow}}, \bibinfo
  {author} {\bibfnamefont {D.~B.}\ \bibnamefont {Sinars}}, \bibinfo {author}
  {\bibfnamefont {K.~D.}\ \bibnamefont {Hahn}}, \bibinfo {author}
  {\bibfnamefont {S.~B.}\ \bibnamefont {Hansen}}, \bibinfo {author}
  {\bibfnamefont {E.~C.}\ \bibnamefont {Harding}}, \bibinfo {author}
  {\bibfnamefont {P.~F.}\ \bibnamefont {Knapp}}, \bibinfo {author}
  {\bibfnamefont {P.~F.}\ \bibnamefont {Schmit}}, \bibinfo {author}
  {\bibfnamefont {C.~A.}\ \bibnamefont {Jennings}}, \bibinfo {author}
  {\bibfnamefont {T.~J.}\ \bibnamefont {Awe}}, \bibinfo {author} {\bibfnamefont
  {M.}~\bibnamefont {Geissel}}, \bibinfo {author} {\bibfnamefont {D.~C.}\
  \bibnamefont {Rovang}}, \bibinfo {author} {\bibfnamefont {G.~A.}\
  \bibnamefont {Chandler}}, \bibinfo {author} {\bibfnamefont {G.~W.}\
  \bibnamefont {Cooper}}, \bibinfo {author} {\bibfnamefont {M.~E.}\
  \bibnamefont {Cuneo}}, \bibinfo {author} {\bibfnamefont {A.~J.}\ \bibnamefont
  {Harvey-Thompson}}, \bibinfo {author} {\bibfnamefont {M.~C.}\ \bibnamefont
  {Herrmann}}, \bibinfo {author} {\bibfnamefont {M.~H.}\ \bibnamefont {Hess}},
  \bibinfo {author} {\bibfnamefont {O.}~\bibnamefont {Johns}}, \bibinfo
  {author} {\bibfnamefont {D.~C.}\ \bibnamefont {Lamppa}}, \bibinfo {author}
  {\bibfnamefont {M.~R.}\ \bibnamefont {Martin}}, \bibinfo {author}
  {\bibfnamefont {R.~D.}\ \bibnamefont {McBride}}, \bibinfo {author}
  {\bibfnamefont {K.~J.}\ \bibnamefont {Peterson}}, \bibinfo {author}
  {\bibfnamefont {J.~L.}\ \bibnamefont {Porter}}, \bibinfo {author}
  {\bibfnamefont {G.~K.}\ \bibnamefont {Robertson}}, \bibinfo {author}
  {\bibfnamefont {G.~A.}\ \bibnamefont {Rochau}}, \bibinfo {author}
  {\bibfnamefont {C.~L.}\ \bibnamefont {Ruiz}}, \bibinfo {author}
  {\bibfnamefont {M.~E.}\ \bibnamefont {Savage}}, \bibinfo {author}
  {\bibfnamefont {I.~C.}\ \bibnamefont {Smith}}, \bibinfo {author}
  {\bibfnamefont {W.~A.}\ \bibnamefont {Stygar}}, \ and\ \bibinfo {author}
  {\bibfnamefont {R.~A.}\ \bibnamefont {Vesey}},\ }\href@noop {} {\bibfield
  {journal} {\bibinfo  {journal} {Phys.\ Rev.\ Lett.}\ }\textbf {\bibinfo
  {volume} {113}},\ \bibinfo {pages} {155003} (\bibinfo {year}
  {2014})}\BibitemShut {NoStop}%
\bibitem [{\citenamefont {Schmit}\ \emph {et~al.}(2014)\citenamefont {Schmit},
  \citenamefont {Knapp}, \citenamefont {Hansen}, \citenamefont {Gomez},
  \citenamefont {Hahn}, \citenamefont {Sinars}, \citenamefont {Peterson},
  \citenamefont {Slutz}, \citenamefont {Sefkow}, \citenamefont {Awe},
  \citenamefont {Harding}, \citenamefont {Jennings}, \citenamefont {Chandler},
  \citenamefont {Cooper}, \citenamefont {Cuneo}, \citenamefont {Geissel},
  \citenamefont {Harvey-Thompson}, \citenamefont {Herrmann}, \citenamefont
  {Hess}, \citenamefont {Johns}, \citenamefont {Lamppa}, \citenamefont
  {Martin}, \citenamefont {McBride}, \citenamefont {Porter}, \citenamefont
  {Robertson}, \citenamefont {Rochau}, \citenamefont {Rovang}, \citenamefont
  {Ruiz}, \citenamefont {Savage}, \citenamefont {Smith}, \citenamefont
  {Stygar},\ and\ \citenamefont {Vesey}}]{schmit14}%
  \BibitemOpen
  \bibfield  {author} {\bibinfo {author} {\bibfnamefont {P.~F.}\ \bibnamefont
  {Schmit}}, \bibinfo {author} {\bibfnamefont {P.~F.}\ \bibnamefont {Knapp}},
  \bibinfo {author} {\bibfnamefont {S.~B.}\ \bibnamefont {Hansen}}, \bibinfo
  {author} {\bibfnamefont {M.~R.}\ \bibnamefont {Gomez}}, \bibinfo {author}
  {\bibfnamefont {K.~D.}\ \bibnamefont {Hahn}}, \bibinfo {author}
  {\bibfnamefont {D.~B.}\ \bibnamefont {Sinars}}, \bibinfo {author}
  {\bibfnamefont {K.~J.}\ \bibnamefont {Peterson}}, \bibinfo {author}
  {\bibfnamefont {S.~A.}\ \bibnamefont {Slutz}}, \bibinfo {author}
  {\bibfnamefont {A.~B.}\ \bibnamefont {Sefkow}}, \bibinfo {author}
  {\bibfnamefont {T.~J.}\ \bibnamefont {Awe}}, \bibinfo {author} {\bibfnamefont
  {E.}~\bibnamefont {Harding}}, \bibinfo {author} {\bibfnamefont {C.~A.}\
  \bibnamefont {Jennings}}, \bibinfo {author} {\bibfnamefont {G.~A.}\
  \bibnamefont {Chandler}}, \bibinfo {author} {\bibfnamefont {G.~W.}\
  \bibnamefont {Cooper}}, \bibinfo {author} {\bibfnamefont {M.~E.}\
  \bibnamefont {Cuneo}}, \bibinfo {author} {\bibfnamefont {M.}~\bibnamefont
  {Geissel}}, \bibinfo {author} {\bibfnamefont {A.~J.}\ \bibnamefont
  {Harvey-Thompson}}, \bibinfo {author} {\bibfnamefont {M.~C.}\ \bibnamefont
  {Herrmann}}, \bibinfo {author} {\bibfnamefont {M.~H.}\ \bibnamefont {Hess}},
  \bibinfo {author} {\bibfnamefont {O.}~\bibnamefont {Johns}}, \bibinfo
  {author} {\bibfnamefont {D.~C.}\ \bibnamefont {Lamppa}}, \bibinfo {author}
  {\bibfnamefont {M.~R.}\ \bibnamefont {Martin}}, \bibinfo {author}
  {\bibfnamefont {R.~D.}\ \bibnamefont {McBride}}, \bibinfo {author}
  {\bibfnamefont {J.~L.}\ \bibnamefont {Porter}}, \bibinfo {author}
  {\bibfnamefont {G.~K.}\ \bibnamefont {Robertson}}, \bibinfo {author}
  {\bibfnamefont {G.~A.}\ \bibnamefont {Rochau}}, \bibinfo {author}
  {\bibfnamefont {D.~C.}\ \bibnamefont {Rovang}}, \bibinfo {author}
  {\bibfnamefont {C.~L.}\ \bibnamefont {Ruiz}}, \bibinfo {author}
  {\bibfnamefont {M.~E.}\ \bibnamefont {Savage}}, \bibinfo {author}
  {\bibfnamefont {I.~C.}\ \bibnamefont {Smith}}, \bibinfo {author}
  {\bibfnamefont {W.~A.}\ \bibnamefont {Stygar}}, \ and\ \bibinfo {author}
  {\bibfnamefont {R.~A.}\ \bibnamefont {Vesey}},\ }\href@noop {} {\bibfield
  {journal} {\bibinfo  {journal} {Phys.\ Rev.\ Lett.}\ }\textbf {\bibinfo
  {volume} {113}},\ \bibinfo {pages} {155004} (\bibinfo {year}
  {2014})}\BibitemShut {NoStop}%
\bibitem [{\citenamefont {Thio}\ \emph {et~al.}(1999)\citenamefont {Thio},
  \citenamefont {Panarella}, \citenamefont {Kirkpatrick}, \citenamefont
  {Knapp}, \citenamefont {Wysocki}, \citenamefont {Parks},\ and\ \citenamefont
  {Schmidt}}]{thio99}%
  \BibitemOpen
  \bibfield  {author} {\bibinfo {author} {\bibfnamefont {Y.~C.~F.}\
  \bibnamefont {Thio}}, \bibinfo {author} {\bibfnamefont {E.}~\bibnamefont
  {Panarella}}, \bibinfo {author} {\bibfnamefont {R.~C.}\ \bibnamefont
  {Kirkpatrick}}, \bibinfo {author} {\bibfnamefont {C.~E.}\ \bibnamefont
  {Knapp}}, \bibinfo {author} {\bibfnamefont {F.}~\bibnamefont {Wysocki}},
  \bibinfo {author} {\bibfnamefont {P.}~\bibnamefont {Parks}}, \ and\ \bibinfo
  {author} {\bibfnamefont {G.}~\bibnamefont {Schmidt}},\ }in\ \href@noop {}
  {\emph {\bibinfo {booktitle} {Current Trends in International Fusion
  Research--Proceedings of the Second International Symposium}}},\ \bibinfo
  {editor} {edited by\ \bibinfo {editor} {\bibfnamefont {E.}~\bibnamefont
  {Panarella}}}\ (\bibinfo  {publisher} {NRC Canada},\ \bibinfo {address}
  {Ottawa},\ \bibinfo {year} {1999})\ p.\ \bibinfo {pages} {113}\BibitemShut
  {NoStop}%
\bibitem [{\citenamefont {Hsu}(2009)}]{hsu09}%
  \BibitemOpen
  \bibfield  {author} {\bibinfo {author} {\bibfnamefont {S.~C.}\ \bibnamefont
  {Hsu}},\ }\href@noop {} {\bibfield  {journal} {\bibinfo  {journal} {J. Fusion
  Energy}\ }\textbf {\bibinfo {volume} {28}},\ \bibinfo {pages} {246} (\bibinfo
  {year} {2009})}\BibitemShut {NoStop}%
\bibitem [{\citenamefont {Hsu}\ \emph {et~al.}(2012{\natexlab{a}})\citenamefont
  {Hsu}, \citenamefont {Awe}, \citenamefont {Brockington}, \citenamefont
  {Case}, \citenamefont {Cassibry}, \citenamefont {Kagan}, \citenamefont
  {Messer}, \citenamefont {Stanic}, \citenamefont {Tang}, \citenamefont
  {Welch},\ and\ \citenamefont {Witherspoon}}]{hsu12}%
  \BibitemOpen
  \bibfield  {author} {\bibinfo {author} {\bibfnamefont {S.~C.}\ \bibnamefont
  {Hsu}}, \bibinfo {author} {\bibfnamefont {T.~J.}\ \bibnamefont {Awe}},
  \bibinfo {author} {\bibfnamefont {S.}~\bibnamefont {Brockington}}, \bibinfo
  {author} {\bibfnamefont {A.}~\bibnamefont {Case}}, \bibinfo {author}
  {\bibfnamefont {J.~T.}\ \bibnamefont {Cassibry}}, \bibinfo {author}
  {\bibfnamefont {G.}~\bibnamefont {Kagan}}, \bibinfo {author} {\bibfnamefont
  {S.~J.}\ \bibnamefont {Messer}}, \bibinfo {author} {\bibfnamefont
  {M.}~\bibnamefont {Stanic}}, \bibinfo {author} {\bibfnamefont
  {X.}~\bibnamefont {Tang}}, \bibinfo {author} {\bibfnamefont {D.~R.}\
  \bibnamefont {Welch}}, \ and\ \bibinfo {author} {\bibfnamefont {F.~D.}\
  \bibnamefont {Witherspoon}},\ }\href@noop {} {\bibfield  {journal} {\bibinfo
  {journal} {IEEE Trans.\ Plasma Sci.}\ }\textbf {\bibinfo {volume} {40}},\
  \bibinfo {pages} {1287} (\bibinfo {year} {2012}{\natexlab{a}})}\BibitemShut
  {NoStop}%
\bibitem [{\citenamefont {Lindemuth}\ and\ \citenamefont
  {Kirkpatrick}(1991)}]{lindemuth1991promise}%
  \BibitemOpen
  \bibfield  {author} {\bibinfo {author} {\bibfnamefont {I.}~\bibnamefont
  {Lindemuth}}\ and\ \bibinfo {author} {\bibfnamefont {R.}~\bibnamefont
  {Kirkpatrick}},\ }\href@noop {} {\bibfield  {journal} {\bibinfo  {journal}
  {Fus.\ Tech.}\ }\textbf {\bibinfo {volume} {20}},\ \bibinfo {pages} {829}
  (\bibinfo {year} {1991})}\BibitemShut {NoStop}%
\bibitem [{\citenamefont {Witherspoon}\ \emph {et~al.}(2009)\citenamefont
  {Witherspoon}, \citenamefont {Case}, \citenamefont {Messer}, \citenamefont
  {\mbox{Bomgardner~II}}, \citenamefont {Phillips}, \citenamefont
  {Brockington},\ and\ \citenamefont {Elton}}]{witherspoon09}%
  \BibitemOpen
  \bibfield  {author} {\bibinfo {author} {\bibfnamefont {F.~D.}\ \bibnamefont
  {Witherspoon}}, \bibinfo {author} {\bibfnamefont {A.}~\bibnamefont {Case}},
  \bibinfo {author} {\bibfnamefont {S.~J.}\ \bibnamefont {Messer}}, \bibinfo
  {author} {\bibfnamefont {R.}~\bibnamefont {\mbox{Bomgardner~II}}}, \bibinfo
  {author} {\bibfnamefont {M.~W.}\ \bibnamefont {Phillips}}, \bibinfo {author}
  {\bibfnamefont {S.}~\bibnamefont {Brockington}}, \ and\ \bibinfo {author}
  {\bibfnamefont {R.}~\bibnamefont {Elton}},\ }\href@noop {} {\bibfield
  {journal} {\bibinfo  {journal} {Rev.\ Sci.\ Instrum.}\ }\textbf {\bibinfo
  {volume} {80}},\ \bibinfo {pages} {083506} (\bibinfo {year}
  {2009})}\BibitemShut {NoStop}%
\bibitem [{\citenamefont {Knapp}\ and\ \citenamefont
  {Kirkpatrick}(2014)}]{knapp14}%
  \BibitemOpen
  \bibfield  {author} {\bibinfo {author} {\bibfnamefont {C.~E.}\ \bibnamefont
  {Knapp}}\ and\ \bibinfo {author} {\bibfnamefont {R.~C.}\ \bibnamefont
  {Kirkpatrick}},\ }\href@noop {} {\bibfield  {journal} {\bibinfo  {journal}
  {Phys.\ Plasmas}\ }\textbf {\bibinfo {volume} {21}},\ \bibinfo {eid} {070701}
  (\bibinfo {year} {2014})}\BibitemShut {NoStop}%
\bibitem [{\citenamefont {Parks}(2008)}]{parks08}%
  \BibitemOpen
  \bibfield  {author} {\bibinfo {author} {\bibfnamefont {P.~B.}\ \bibnamefont
  {Parks}},\ }\href@noop {} {\bibfield  {journal} {\bibinfo  {journal} {Phys.\
  Plasmas}\ }\textbf {\bibinfo {volume} {15}},\ \bibinfo {pages} {062506}
  (\bibinfo {year} {2008})}\BibitemShut {NoStop}%
\bibitem [{\citenamefont {Cassibry}\ \emph {et~al.}(2009)\citenamefont
  {Cassibry}, \citenamefont {Cortez}, \citenamefont {Hsu},\ and\ \citenamefont
  {Witherspoon}}]{cassibry09}%
  \BibitemOpen
  \bibfield  {author} {\bibinfo {author} {\bibfnamefont {J.~T.}\ \bibnamefont
  {Cassibry}}, \bibinfo {author} {\bibfnamefont {R.~J.}\ \bibnamefont
  {Cortez}}, \bibinfo {author} {\bibfnamefont {S.~C.}\ \bibnamefont {Hsu}}, \
  and\ \bibinfo {author} {\bibfnamefont {F.~D.}\ \bibnamefont {Witherspoon}},\
  }\href@noop {} {\bibfield  {journal} {\bibinfo  {journal} {Phys.\ Plasmas}\
  }\textbf {\bibinfo {volume} {16}},\ \bibinfo {pages} {112707} (\bibinfo
  {year} {2009})}\BibitemShut {NoStop}%
\bibitem [{\citenamefont {Awe}\ \emph {et~al.}(2011)\citenamefont {Awe},
  \citenamefont {Adams}, \citenamefont {Davis}, \citenamefont {Hanna},
  \citenamefont {Hsu},\ and\ \citenamefont {Cassibry}}]{awe11}%
  \BibitemOpen
  \bibfield  {author} {\bibinfo {author} {\bibfnamefont {T.~J.}\ \bibnamefont
  {Awe}}, \bibinfo {author} {\bibfnamefont {C.~S.}\ \bibnamefont {Adams}},
  \bibinfo {author} {\bibfnamefont {J.~S.}\ \bibnamefont {Davis}}, \bibinfo
  {author} {\bibfnamefont {D.~S.}\ \bibnamefont {Hanna}}, \bibinfo {author}
  {\bibfnamefont {S.~C.}\ \bibnamefont {Hsu}}, \ and\ \bibinfo {author}
  {\bibfnamefont {J.~T.}\ \bibnamefont {Cassibry}},\ }\href@noop {} {\bibfield
  {journal} {\bibinfo  {journal} {Phys.\ Plasmas}\ }\textbf {\bibinfo {volume}
  {18}},\ \bibinfo {pages} {072705} (\bibinfo {year} {2011})}\BibitemShut
  {NoStop}%
\bibitem [{\citenamefont {Davis}\ \emph {et~al.}(2012)\citenamefont {Davis},
  \citenamefont {Hsu}, \citenamefont {Golovkin}, \citenamefont {MacFarlane},\
  and\ \citenamefont {Cassibry}}]{davis12}%
  \BibitemOpen
  \bibfield  {author} {\bibinfo {author} {\bibfnamefont {J.~S.}\ \bibnamefont
  {Davis}}, \bibinfo {author} {\bibfnamefont {S.~C.}\ \bibnamefont {Hsu}},
  \bibinfo {author} {\bibfnamefont {I.~E.}\ \bibnamefont {Golovkin}}, \bibinfo
  {author} {\bibfnamefont {J.~J.}\ \bibnamefont {MacFarlane}}, \ and\ \bibinfo
  {author} {\bibfnamefont {J.~T.}\ \bibnamefont {Cassibry}},\ }\href@noop {}
  {\bibfield  {journal} {\bibinfo  {journal} {Phys.\ Plasmas}\ }\textbf
  {\bibinfo {volume} {19}},\ \bibinfo {pages} {102701} (\bibinfo {year}
  {2012})}\BibitemShut {NoStop}%
\bibitem [{\citenamefont {Samulyak}, \citenamefont {Parks},\ and\ \citenamefont
  {Wu}(2010)}]{samulyak10}%
  \BibitemOpen
  \bibfield  {author} {\bibinfo {author} {\bibfnamefont {R.}~\bibnamefont
  {Samulyak}}, \bibinfo {author} {\bibfnamefont {P.}~\bibnamefont {Parks}}, \
  and\ \bibinfo {author} {\bibfnamefont {L.}~\bibnamefont {Wu}},\ }\href@noop
  {} {\bibfield  {journal} {\bibinfo  {journal} {Phys.\ Plasmas}\ }\textbf
  {\bibinfo {volume} {17}},\ \bibinfo {pages} {092702} (\bibinfo {year}
  {2010})}\BibitemShut {NoStop}%
\bibitem [{\citenamefont {Santarius}(2012)}]{santarius12}%
  \BibitemOpen
  \bibfield  {author} {\bibinfo {author} {\bibfnamefont {J.~F.}\ \bibnamefont
  {Santarius}},\ }\href@noop {} {\bibfield  {journal} {\bibinfo  {journal}
  {Phys.\ Plasmas}\ }\textbf {\bibinfo {volume} {19}},\ \bibinfo {pages}
  {072705} (\bibinfo {year} {2012})}\BibitemShut {NoStop}%
\bibitem [{\citenamefont {Kim}\ \emph {et~al.}(2012)\citenamefont {Kim},
  \citenamefont {Samulyak}, \citenamefont {Zhang},\ and\ \citenamefont
  {Parks}}]{kim12}%
  \BibitemOpen
  \bibfield  {author} {\bibinfo {author} {\bibfnamefont {H.}~\bibnamefont
  {Kim}}, \bibinfo {author} {\bibfnamefont {R.}~\bibnamefont {Samulyak}},
  \bibinfo {author} {\bibfnamefont {L.}~\bibnamefont {Zhang}}, \ and\ \bibinfo
  {author} {\bibfnamefont {P.}~\bibnamefont {Parks}},\ }\href@noop {}
  {\bibfield  {journal} {\bibinfo  {journal} {Phys.\ Plasmas}\ }\textbf
  {\bibinfo {volume} {19}},\ \bibinfo {pages} {082711} (\bibinfo {year}
  {2012})}\BibitemShut {NoStop}%
\bibitem [{\citenamefont {Cassibry}\ \emph {et~al.}(2012)\citenamefont
  {Cassibry}, \citenamefont {Stanic}, \citenamefont {Hsu}, \citenamefont
  {Witherspoon},\ and\ \citenamefont {Abarzhi}}]{cassibry12}%
  \BibitemOpen
  \bibfield  {author} {\bibinfo {author} {\bibfnamefont {J.~T.}\ \bibnamefont
  {Cassibry}}, \bibinfo {author} {\bibfnamefont {M.}~\bibnamefont {Stanic}},
  \bibinfo {author} {\bibfnamefont {S.~C.}\ \bibnamefont {Hsu}}, \bibinfo
  {author} {\bibfnamefont {F.~D.}\ \bibnamefont {Witherspoon}}, \ and\ \bibinfo
  {author} {\bibfnamefont {S.~I.}\ \bibnamefont {Abarzhi}},\ }\href@noop {}
  {\bibfield  {journal} {\bibinfo  {journal} {Phys.\ Plasmas}\ }\textbf
  {\bibinfo {volume} {19}},\ \bibinfo {pages} {052702} (\bibinfo {year}
  {2012})}\BibitemShut {NoStop}%
\bibitem [{\citenamefont {Kim}\ \emph {et~al.}(2013)\citenamefont {Kim},
  \citenamefont {Zhang}, \citenamefont {Samulyak},\ and\ \citenamefont
  {Parks}}]{kim13}%
  \BibitemOpen
  \bibfield  {author} {\bibinfo {author} {\bibfnamefont {H.}~\bibnamefont
  {Kim}}, \bibinfo {author} {\bibfnamefont {L.}~\bibnamefont {Zhang}}, \bibinfo
  {author} {\bibfnamefont {R.}~\bibnamefont {Samulyak}}, \ and\ \bibinfo
  {author} {\bibfnamefont {P.}~\bibnamefont {Parks}},\ }\href@noop {}
  {\bibfield  {journal} {\bibinfo  {journal} {Phys.\ Plasmas}\ }\textbf
  {\bibinfo {volume} {20}},\ \bibinfo {eid} {022704} (\bibinfo {year}
  {2013})}\BibitemShut {NoStop}%
\bibitem [{\citenamefont {Cassibry}, \citenamefont {Stanic},\ and\
  \citenamefont {Hsu}(2013)}]{cassibry13}%
  \BibitemOpen
  \bibfield  {author} {\bibinfo {author} {\bibfnamefont {J.~T.}\ \bibnamefont
  {Cassibry}}, \bibinfo {author} {\bibfnamefont {M.}~\bibnamefont {Stanic}}, \
  and\ \bibinfo {author} {\bibfnamefont {S.~C.}\ \bibnamefont {Hsu}},\
  }\href@noop {} {\bibfield  {journal} {\bibinfo  {journal} {Phys.\ Plasmas}\
  }\textbf {\bibinfo {volume} {20}},\ \bibinfo {eid} {032706} (\bibinfo {year}
  {2013})}\BibitemShut {NoStop}%
\bibitem [{\citenamefont {Case}\ \emph {et~al.}(2010)\citenamefont {Case},
  \citenamefont {Messer}, \citenamefont {Bomgardner},\ and\ \citenamefont
  {Witherspoon}}]{case10}%
  \BibitemOpen
  \bibfield  {author} {\bibinfo {author} {\bibfnamefont {A.}~\bibnamefont
  {Case}}, \bibinfo {author} {\bibfnamefont {S.}~\bibnamefont {Messer}},
  \bibinfo {author} {\bibfnamefont {R.}~\bibnamefont {Bomgardner}}, \ and\
  \bibinfo {author} {\bibfnamefont {F.~D.}\ \bibnamefont {Witherspoon}},\
  }\href@noop {} {\bibfield  {journal} {\bibinfo  {journal} {Phys.\ Plasmas}\
  }\textbf {\bibinfo {volume} {17}},\ \bibinfo {pages} {053503} (\bibinfo
  {year} {2010})}\BibitemShut {NoStop}%
\bibitem [{\citenamefont {Hsu}\ \emph {et~al.}(2012{\natexlab{b}})\citenamefont
  {Hsu}, \citenamefont {Merritt}, \citenamefont {Moser}, \citenamefont {Awe},
  \citenamefont {Brockington}, \citenamefont {Davis}, \citenamefont {Adams},
  \citenamefont {Case}, \citenamefont {Cassibry}, \citenamefont {Dunn},
  \citenamefont {Gilmore}, \citenamefont {Lynn}, \citenamefont {Messer},\ and\
  \citenamefont {Witherspoon}}]{hsu12pop}%
  \BibitemOpen
  \bibfield  {author} {\bibinfo {author} {\bibfnamefont {S.~C.}\ \bibnamefont
  {Hsu}}, \bibinfo {author} {\bibfnamefont {E.~C.}\ \bibnamefont {Merritt}},
  \bibinfo {author} {\bibfnamefont {A.~L.}\ \bibnamefont {Moser}}, \bibinfo
  {author} {\bibfnamefont {T.~J.}\ \bibnamefont {Awe}}, \bibinfo {author}
  {\bibfnamefont {S.~J.~E.}\ \bibnamefont {Brockington}}, \bibinfo {author}
  {\bibfnamefont {J.~S.}\ \bibnamefont {Davis}}, \bibinfo {author}
  {\bibfnamefont {C.~S.}\ \bibnamefont {Adams}}, \bibinfo {author}
  {\bibfnamefont {A.}~\bibnamefont {Case}}, \bibinfo {author} {\bibfnamefont
  {J.~T.}\ \bibnamefont {Cassibry}}, \bibinfo {author} {\bibfnamefont {J.~P.}\
  \bibnamefont {Dunn}}, \bibinfo {author} {\bibfnamefont {M.~A.}\ \bibnamefont
  {Gilmore}}, \bibinfo {author} {\bibfnamefont {A.~G.}\ \bibnamefont {Lynn}},
  \bibinfo {author} {\bibfnamefont {S.~J.}\ \bibnamefont {Messer}}, \ and\
  \bibinfo {author} {\bibfnamefont {F.~D.}\ \bibnamefont {Witherspoon}},\
  }\href@noop {} {\bibfield  {journal} {\bibinfo  {journal} {Phys.\ Plasmas}\
  }\textbf {\bibinfo {volume} {19}},\ \bibinfo {eid} {123514} (\bibinfo {year}
  {2012}{\natexlab{b}})}\BibitemShut {NoStop}%
\bibitem [{\citenamefont {Merritt}\ \emph {et~al.}(2013)\citenamefont
  {Merritt}, \citenamefont {Moser}, \citenamefont {Hsu}, \citenamefont
  {Loverich},\ and\ \citenamefont {Gilmore}}]{merritt13}%
  \BibitemOpen
  \bibfield  {author} {\bibinfo {author} {\bibfnamefont {E.~C.}\ \bibnamefont
  {Merritt}}, \bibinfo {author} {\bibfnamefont {A.~L.}\ \bibnamefont {Moser}},
  \bibinfo {author} {\bibfnamefont {S.~C.}\ \bibnamefont {Hsu}}, \bibinfo
  {author} {\bibfnamefont {J.}~\bibnamefont {Loverich}}, \ and\ \bibinfo
  {author} {\bibfnamefont {M.}~\bibnamefont {Gilmore}},\ }\href@noop {}
  {\bibfield  {journal} {\bibinfo  {journal} {Phys.\ Rev.\ Lett.}\ }\textbf
  {\bibinfo {volume} {111}},\ \bibinfo {pages} {085003} (\bibinfo {year}
  {2013})}\BibitemShut {NoStop}%
\bibitem [{\citenamefont {Merritt}\ \emph {et~al.}(2014)\citenamefont
  {Merritt}, \citenamefont {Moser}, \citenamefont {Hsu}, \citenamefont {Adams},
  \citenamefont {Dunn}, \citenamefont {Miguel~Holgado},\ and\ \citenamefont
  {Gilmore}}]{merritt14}%
  \BibitemOpen
  \bibfield  {author} {\bibinfo {author} {\bibfnamefont {E.~C.}\ \bibnamefont
  {Merritt}}, \bibinfo {author} {\bibfnamefont {A.~L.}\ \bibnamefont {Moser}},
  \bibinfo {author} {\bibfnamefont {S.~C.}\ \bibnamefont {Hsu}}, \bibinfo
  {author} {\bibfnamefont {C.~S.}\ \bibnamefont {Adams}}, \bibinfo {author}
  {\bibfnamefont {J.~P.}\ \bibnamefont {Dunn}}, \bibinfo {author}
  {\bibfnamefont {A.}~\bibnamefont {Miguel~Holgado}}, \ and\ \bibinfo {author}
  {\bibfnamefont {M.~A.}\ \bibnamefont {Gilmore}},\ }\href@noop {} {\bibfield
  {journal} {\bibinfo  {journal} {Phys.\ Plasmas}\ }\textbf {\bibinfo {volume}
  {21}},\ \bibinfo {eid} {055703} (\bibinfo {year} {2014})}\BibitemShut
  {NoStop}%
\bibitem [{\citenamefont {Case}\ \emph {et~al.}(2013)\citenamefont {Case},
  \citenamefont {Messer}, \citenamefont {Brockington}, \citenamefont {Wu},
  \citenamefont {Witherspoon},\ and\ \citenamefont {Elton}}]{case13}%
  \BibitemOpen
  \bibfield  {author} {\bibinfo {author} {\bibfnamefont {A.}~\bibnamefont
  {Case}}, \bibinfo {author} {\bibfnamefont {S.}~\bibnamefont {Messer}},
  \bibinfo {author} {\bibfnamefont {S.}~\bibnamefont {Brockington}}, \bibinfo
  {author} {\bibfnamefont {L.}~\bibnamefont {Wu}}, \bibinfo {author}
  {\bibfnamefont {F.~D.}\ \bibnamefont {Witherspoon}}, \ and\ \bibinfo {author}
  {\bibfnamefont {R.}~\bibnamefont {Elton}},\ }\href@noop {} {\bibfield
  {journal} {\bibinfo  {journal} {Phys.\ Plasmas}\ }\textbf {\bibinfo {volume}
  {20}},\ \bibinfo {eid} {012704} (\bibinfo {year} {2013})}\BibitemShut
  {NoStop}%
\bibitem [{\citenamefont {Messer}\ \emph {et~al.}(2013)\citenamefont {Messer},
  \citenamefont {Case}, \citenamefont {Wu}, \citenamefont {Brockington},\ and\
  \citenamefont {Witherspoon}}]{messer13}%
  \BibitemOpen
  \bibfield  {author} {\bibinfo {author} {\bibfnamefont {S.}~\bibnamefont
  {Messer}}, \bibinfo {author} {\bibfnamefont {A.}~\bibnamefont {Case}},
  \bibinfo {author} {\bibfnamefont {L.}~\bibnamefont {Wu}}, \bibinfo {author}
  {\bibfnamefont {S.}~\bibnamefont {Brockington}}, \ and\ \bibinfo {author}
  {\bibfnamefont {F.~D.}\ \bibnamefont {Witherspoon}},\ }\href@noop {}
  {\bibfield  {journal} {\bibinfo  {journal} {Phys.\ Plasmas}\ }\textbf
  {\bibinfo {volume} {20}},\ \bibinfo {eid} {032306} (\bibinfo {year}
  {2013})}\BibitemShut {NoStop}%
\bibitem [{\citenamefont {McBride}\ and\ \citenamefont
  {Slutz}(2015)}]{mcbride15}%
  \BibitemOpen
  \bibfield  {author} {\bibinfo {author} {\bibfnamefont {R.~D.}\ \bibnamefont
  {McBride}}\ and\ \bibinfo {author} {\bibfnamefont {S.~A.}\ \bibnamefont
  {Slutz}},\ }\href@noop {} {\bibfield  {journal} {\bibinfo  {journal} {Phys.\
  Plasmas}\ }\textbf {\bibinfo {volume} {22}},\ \bibinfo {eid} {052708}
  (\bibinfo {year} {2015})}\BibitemShut {NoStop}%
\bibitem [{\citenamefont {McBride}\ \emph {et~al.}(2016)\citenamefont
  {McBride}, \citenamefont {Slutz}, \citenamefont {Vesey}, \citenamefont
  {Gomez}, \citenamefont {Sefkow}, \citenamefont {Hansen}, \citenamefont
  {Knapp}, \citenamefont {Schmit}, \citenamefont {Geissel}, \citenamefont
  {Harvey-Thompson}, \citenamefont {Jennings}, \citenamefont {Harding},
  \citenamefont {Awe}, \citenamefont {Rovang}, \citenamefont {Hahn},
  \citenamefont {Martin}, \citenamefont {Cochrane}, \citenamefont {Peterson},
  \citenamefont {Rochau}, \citenamefont {Porter}, \citenamefont {Stygar},
  \citenamefont {Campbell}, \citenamefont {Nakhleh}, \citenamefont {Herrmann},
  \citenamefont {Cuneo},\ and\ \citenamefont {Sinars}}]{mcbride16}%
  \BibitemOpen
  \bibfield  {author} {\bibinfo {author} {\bibfnamefont {R.~D.}\ \bibnamefont
  {McBride}}, \bibinfo {author} {\bibfnamefont {S.~A.}\ \bibnamefont {Slutz}},
  \bibinfo {author} {\bibfnamefont {R.~A.}\ \bibnamefont {Vesey}}, \bibinfo
  {author} {\bibfnamefont {M.~R.}\ \bibnamefont {Gomez}}, \bibinfo {author}
  {\bibfnamefont {A.~B.}\ \bibnamefont {Sefkow}}, \bibinfo {author}
  {\bibfnamefont {S.~B.}\ \bibnamefont {Hansen}}, \bibinfo {author}
  {\bibfnamefont {P.~F.}\ \bibnamefont {Knapp}}, \bibinfo {author}
  {\bibfnamefont {P.~F.}\ \bibnamefont {Schmit}}, \bibinfo {author}
  {\bibfnamefont {M.}~\bibnamefont {Geissel}}, \bibinfo {author} {\bibfnamefont
  {A.~J.}\ \bibnamefont {Harvey-Thompson}}, \bibinfo {author} {\bibfnamefont
  {C.~A.}\ \bibnamefont {Jennings}}, \bibinfo {author} {\bibfnamefont {E.~C.}\
  \bibnamefont {Harding}}, \bibinfo {author} {\bibfnamefont {T.~J.}\
  \bibnamefont {Awe}}, \bibinfo {author} {\bibfnamefont {D.~C.}\ \bibnamefont
  {Rovang}}, \bibinfo {author} {\bibfnamefont {K.~D.}\ \bibnamefont {Hahn}},
  \bibinfo {author} {\bibfnamefont {M.~R.}\ \bibnamefont {Martin}}, \bibinfo
  {author} {\bibfnamefont {K.~R.}\ \bibnamefont {Cochrane}}, \bibinfo {author}
  {\bibfnamefont {K.~J.}\ \bibnamefont {Peterson}}, \bibinfo {author}
  {\bibfnamefont {G.~A.}\ \bibnamefont {Rochau}}, \bibinfo {author}
  {\bibfnamefont {J.~L.}\ \bibnamefont {Porter}}, \bibinfo {author}
  {\bibfnamefont {W.~A.}\ \bibnamefont {Stygar}}, \bibinfo {author}
  {\bibfnamefont {E.~M.}\ \bibnamefont {Campbell}}, \bibinfo {author}
  {\bibfnamefont {C.~W.}\ \bibnamefont {Nakhleh}}, \bibinfo {author}
  {\bibfnamefont {M.~C.}\ \bibnamefont {Herrmann}}, \bibinfo {author}
  {\bibfnamefont {M.~E.}\ \bibnamefont {Cuneo}}, \ and\ \bibinfo {author}
  {\bibfnamefont {D.~B.}\ \bibnamefont {Sinars}},\ }\href@noop {} {\bibfield
  {journal} {\bibinfo  {journal} {Phys.\ Plasmas}\ }\textbf {\bibinfo {volume}
  {23}},\ \bibinfo {eid} {012705} (\bibinfo {year} {2016})}\BibitemShut
  {NoStop}%
\bibitem [{cod()}]{code}%
  \BibitemOpen
  \href@noop {} {}\bibinfo {note} {See supplemental materials to access the
  numerical code based on the semi-analytic model presented in this
  work.}\BibitemShut {Stop}%
\bibitem [{\citenamefont {Welch}\ \emph {et~al.}(2012)\citenamefont {Welch},
  \citenamefont {Genoni}, \citenamefont {Thoma}, \citenamefont {Bruner},
  \citenamefont {Rose},\ and\ \citenamefont {Hsu}}]{welch2012simulations}%
  \BibitemOpen
  \bibfield  {author} {\bibinfo {author} {\bibfnamefont {D.}~\bibnamefont
  {Welch}}, \bibinfo {author} {\bibfnamefont {T.}~\bibnamefont {Genoni}},
  \bibinfo {author} {\bibfnamefont {C.}~\bibnamefont {Thoma}}, \bibinfo
  {author} {\bibfnamefont {N.}~\bibnamefont {Bruner}}, \bibinfo {author}
  {\bibfnamefont {D.}~\bibnamefont {Rose}}, \ and\ \bibinfo {author}
  {\bibfnamefont {S.}~\bibnamefont {Hsu}},\ }\href@noop {} {\bibfield
  {journal} {\bibinfo  {journal} {Phys.\ Rev.\ Lett.}\ }\textbf {\bibinfo
  {volume} {109}},\ \bibinfo {pages} {225002} (\bibinfo {year}
  {2012})}\BibitemShut {NoStop}%
\bibitem [{\citenamefont {Welch}\ \emph {et~al.}(2014)\citenamefont {Welch},
  \citenamefont {Genoni}, \citenamefont {Thoma}, \citenamefont {Rose},\ and\
  \citenamefont {Hsu}}]{welch2014particle}%
  \BibitemOpen
  \bibfield  {author} {\bibinfo {author} {\bibfnamefont {D.}~\bibnamefont
  {Welch}}, \bibinfo {author} {\bibfnamefont {T.}~\bibnamefont {Genoni}},
  \bibinfo {author} {\bibfnamefont {C.}~\bibnamefont {Thoma}}, \bibinfo
  {author} {\bibfnamefont {D.}~\bibnamefont {Rose}}, \ and\ \bibinfo {author}
  {\bibfnamefont {S.}~\bibnamefont {Hsu}},\ }\href@noop {} {\bibfield
  {journal} {\bibinfo  {journal} {Phys.\ Plasmas}\ }\textbf {\bibinfo {volume}
  {21}},\ \bibinfo {pages} {032704} (\bibinfo {year} {2014})}\BibitemShut
  {NoStop}%
\bibitem [{\citenamefont {Kirkpatrick}(1979)}]{kirkpatrick1979overview}%
  \BibitemOpen
  \bibfield  {author} {\bibinfo {author} {\bibfnamefont {R.}~\bibnamefont
  {Kirkpatrick}},\ }\href@noop {} {\bibfield  {journal} {\bibinfo  {journal}
  {Nucl.\ Fusion}\ }\textbf {\bibinfo {volume} {19}},\ \bibinfo {pages} {69}
  (\bibinfo {year} {1979})}\BibitemShut {NoStop}%
\bibitem [{\citenamefont {MacFarlane}, \citenamefont {Golovkin},\ and\
  \citenamefont {Woodruff}(2006)}]{macfarlane06}%
  \BibitemOpen
  \bibfield  {author} {\bibinfo {author} {\bibfnamefont {J.~J.}\ \bibnamefont
  {MacFarlane}}, \bibinfo {author} {\bibfnamefont {I.~E.}\ \bibnamefont
  {Golovkin}}, \ and\ \bibinfo {author} {\bibfnamefont {P.~R.}\ \bibnamefont
  {Woodruff}},\ }\href@noop {} {\bibfield  {journal} {\bibinfo  {journal} {J.\
  Quant.\ Spect.\ Rad.\ Transfer}\ }\textbf {\bibinfo {volume} {99}},\ \bibinfo
  {pages} {381} (\bibinfo {year} {2006})}\BibitemShut {NoStop}%
\bibitem [{\citenamefont {Epperlein}\ and\ \citenamefont
  {Haines}(1986)}]{epperlein1986plasma}%
  \BibitemOpen
  \bibfield  {author} {\bibinfo {author} {\bibfnamefont {E.}~\bibnamefont
  {Epperlein}}\ and\ \bibinfo {author} {\bibfnamefont {M.}~\bibnamefont
  {Haines}},\ }\href@noop {} {\bibfield  {journal} {\bibinfo  {journal} {Phys.\
  Fluids}\ }\textbf {\bibinfo {volume} {29}},\ \bibinfo {pages} {1029}
  (\bibinfo {year} {1986})}\BibitemShut {NoStop}%
\bibitem [{\citenamefont {Velikovich}, \citenamefont {Giuliani},\ and\
  \citenamefont {Zalesak}(2015)}]{velikovich2015magnetic}%
  \BibitemOpen
  \bibfield  {author} {\bibinfo {author} {\bibfnamefont {A.}~\bibnamefont
  {Velikovich}}, \bibinfo {author} {\bibfnamefont {J.}~\bibnamefont
  {Giuliani}}, \ and\ \bibinfo {author} {\bibfnamefont {S.}~\bibnamefont
  {Zalesak}},\ }\href@noop {} {\bibfield  {journal} {\bibinfo  {journal}
  {Phys.\ Plasmas}\ }\textbf {\bibinfo {volume} {22}},\ \bibinfo {pages}
  {042702} (\bibinfo {year} {2015})}\BibitemShut {NoStop}%
\bibitem [{\citenamefont {Braginskii}(1965)}]{braginskii1965transport}%
  \BibitemOpen
  \bibfield  {author} {\bibinfo {author} {\bibfnamefont {S.}~\bibnamefont
  {Braginskii}},\ }\href@noop {} {\bibfield  {journal} {\bibinfo  {journal}
  {Rev.\ Plasma Phys.}\ }\textbf {\bibinfo {volume} {1}},\ \bibinfo {pages}
  {205} (\bibinfo {year} {1965})}\BibitemShut {NoStop}%
\bibitem [{\citenamefont {Epperlein}(1990)}]{epperlein1990kinetic}%
  \BibitemOpen
  \bibfield  {author} {\bibinfo {author} {\bibfnamefont {E.~M.}\ \bibnamefont
  {Epperlein}},\ }\href@noop {} {\bibfield  {journal} {\bibinfo  {journal}
  {Physical review letters}\ }\textbf {\bibinfo {volume} {65}},\ \bibinfo
  {pages} {2145} (\bibinfo {year} {1990})}\BibitemShut {NoStop}%
\bibitem [{\citenamefont {Turner}\ and\ \citenamefont
  {Stone}(2001)}]{turner2001module}%
  \BibitemOpen
  \bibfield  {author} {\bibinfo {author} {\bibfnamefont {N.}~\bibnamefont
  {Turner}}\ and\ \bibinfo {author} {\bibfnamefont {J.}~\bibnamefont {Stone}},\
  }\href@noop {} {\bibfield  {journal} {\bibinfo  {journal} {The Astrophysical
  Journal Supplement Series}\ }\textbf {\bibinfo {volume} {135}},\ \bibinfo
  {pages} {95} (\bibinfo {year} {2001})}\BibitemShut {NoStop}%
\bibitem [{nrl()}]{nrl-formulary}%
  \BibitemOpen
  \href@noop {} {}\bibinfo {note} {J. D. Huba, {\em NRL Plasma Formulary},
  2016.}\BibitemShut {Stop}%
\bibitem [{\citenamefont {Carroll}\ and\ \citenamefont
  {Ostlie}(2006)}]{carroll2006introduction}%
  \BibitemOpen
  \bibfield  {author} {\bibinfo {author} {\bibfnamefont {B.~W.}\ \bibnamefont
  {Carroll}}\ and\ \bibinfo {author} {\bibfnamefont {D.~A.}\ \bibnamefont
  {Ostlie}},\ }\href@noop {} {\emph {\bibinfo {title} {An introduction to
  modern astrophysics and cosmology}}},\ Vol.~\bibinfo {volume} {1}\ (\bibinfo
  {year} {2006})\BibitemShut {NoStop}%
\bibitem [{\citenamefont {Bosch}\ and\ \citenamefont
  {Hale}(1992)}]{bosch1992improved}%
  \BibitemOpen
  \bibfield  {author} {\bibinfo {author} {\bibfnamefont {H.-S.}\ \bibnamefont
  {Bosch}}\ and\ \bibinfo {author} {\bibfnamefont {G.}~\bibnamefont {Hale}},\
  }\href@noop {} {\bibfield  {journal} {\bibinfo  {journal} {Nucl.\ Fusion}\
  }\textbf {\bibinfo {volume} {32}},\ \bibinfo {pages} {611} (\bibinfo {year}
  {1992})}\BibitemShut {NoStop}%
\bibitem [{\citenamefont {Fraley}\ \emph {et~al.}(1974)\citenamefont {Fraley},
  \citenamefont {Linnebur}, \citenamefont {Mason},\ and\ \citenamefont
  {Morse}}]{fraley1974thermonuclear}%
  \BibitemOpen
  \bibfield  {author} {\bibinfo {author} {\bibfnamefont {G.}~\bibnamefont
  {Fraley}}, \bibinfo {author} {\bibfnamefont {E.}~\bibnamefont {Linnebur}},
  \bibinfo {author} {\bibfnamefont {R.}~\bibnamefont {Mason}}, \ and\ \bibinfo
  {author} {\bibfnamefont {R.}~\bibnamefont {Morse}},\ }\href@noop {}
  {\bibfield  {journal} {\bibinfo  {journal} {Phys.\ Fluids}\ }\textbf
  {\bibinfo {volume} {17}},\ \bibinfo {pages} {474} (\bibinfo {year}
  {1974})}\BibitemShut {NoStop}%
\end{thebibliography}
%

\end{document}